\documentclass[twocolumn,english,aps,superscriptaddress,prl,floatfix,longbibliography]{revtex4-2}
\usepackage[latin9]{inputenc}
\usepackage{color}
\usepackage[dvipsnames]{xcolor}
\usepackage{babel}
\usepackage{amsmath}
\usepackage{amssymb}
\usepackage{graphicx}
\usepackage{esint}
\usepackage[unicode=true,pdfusetitle,
 bookmarks=true,bookmarksnumbered=false,bookmarksopen=false,
 breaklinks=true,pdfborder={0 0 0},pdfborderstyle={},backref=false,colorlinks=true]
 {hyperref}
\hypersetup{
 linkcolor=blue,citecolor=blue,urlcolor=blue}

\makeatletter

\usepackage{color}
\usepackage{babel}

\usepackage{color}
\usepackage{dsfont}   
\usepackage{braket}

\newcommand{\rucl}{$\alpha$-RuCl$_3$}
\newcommand{\hlio}{H$_3$LiIr$_2$O$_6$}
\newcommand{\kp}{K_{3}^{\prime}}

\makeatother

\begin{document}
\title{Disorder, Low-Energy Excitations, and Topology in the Kitaev Spin Liquid }
\author{Vitor Dantas}
\affiliation{Instituto de F{\'i}sica de S\~ao Carlos, Universidade de S\~ao Paulo,
C.P. 369, S\~ao Carlos, SP, 13560-970, Brazil}
\author{Eric C. Andrade}
\affiliation{Instituto de F{\'i}sica de S\~ao Carlos, Universidade de S\~ao Paulo,
C.P. 369, S\~ao Carlos, SP, 13560-970, Brazil}
\begin{abstract}
The Kitaev model is a fascinating example of an exactly solvable model displaying a spin-liquid ground state in two dimensions. However, deviations from the original Kitaev model are expected to appear in real materials. In this Letter, we investigate the fate of Kitaev's spin liquid in the presence of disorder ---bond defects or vacancies--- for an extended version of the model. Considering static flux backgrounds, we observe a power-law divergence in the low-energy limit of the density of states with a nonuniversal exponent. We link this power-law distribution of energy scales to weakly coupled droplets inside the bulk,  in an uncanny similarity to the Griffiths phase often present in the vicinity of disordered quantum phase transitions.  If time-reversal symmetry is broken, we find that power-law singularities are tied to the destruction of the topological phase of the Kitaev model in the presence of bond disorder alone.  However,  there is a transition from this topologically trivial phase with power-law singularities to a topologically nontrivial one for weak to moderate site dilution.  Therefore,  diluted Kitaev materials are potential candidates to host Kitaev's chiral spin-liquid phase.
\end{abstract}
\date{\today}

\maketitle
\emph{Introduction.}-- Over the past decades, strong spin-orbit coupling
has been recognized as a key ingredient in stabilizing unconventional
phases in correlated materials \citep{khaliullin05,witczak-krempa14,nussinov15,rau16}.
For $4d$ and $5d$ Mott insulators,  for instance,  there is great interest in the
Kitaev materials, which are systems hosting dominant Ising-like bond-dependent interactions
for local effective moments $j_{{\rm eff}}=1/2$ in stacked honeycomb
planes \citep{kitaev06,jackeli09,chaloupka10,chaloupka13,winter17,trebst17,hermanns18,takagi19}.
If these bond-depended interactions have similar strength,  Kitaev exactly established the existence of a quantum spin liquid \citep{balents10,savary17b,broholm20} of gapless Majorana fermions
moving in a static $\mathbb{Z}_{2}$ flux background.  Remarkably,  an infinitesimally small external magnetic field generates chiral Majorana edge modes with half-quantized thermal Hall conductance \citep{kitaev06,  hickey19}.

A promising Kitaev material is \rucl ~\cite{plumb14,sears15,banerjee16},
which displays long-ranged magnetic order at low $T$, suggesting
further magnetic interactions beyond Kitaev's \citep{janssen17}.
The magnetic order is suppressed by an external magnetic field \citep{sears17,wolter17,kasahara18a,janssen19,balz2019,gass2020,consoli20},
and it is replaced by an intermediary phase ---distinct from the high-field
polarized state--- which exhibits a half-quantized thermal Hall conductance
\citep{kasahara18b,yokoi21}. 

Another putative Kitaev material is \hlio ~\cite{kitagawa18}, which
shows no magnetic order down to $50$ mK, making it a prominent candidate
to realize Kitaev's spin-liquid phase. However, the experimental observations
are at odds with the thermodynamic behavior of the clean Kitaev model
\citep{nasu14,nasu15,feng20}: (i) the specific heat diverges at low $T$
as $C/T\propto T^{-1/2}$; (ii) the uniform magnetic susceptibility
shows a similar, albeit milder,  divergence $\chi\sim T^{-1/2}$; (iii) the $1/T_{1}$
NMR spin-relaxation rate has a nonvanishing contribution down to
low $T$, and the Knight shift is almost flat in this region.
All these results point to an appreciable amount of low-energy excitations.

This work shows that the experimental observations in \hlio \ \ can be understood within Kitaev's model if one considers the
presence of defects. Microscopic sources of the disorder include stacking
faults \citep{freitas21} and the random position of the H ions. To
study the effects of uncorrelated quenched disorder in this model in a controlled
fashion,  we address the role of bond disorder and site
dilution (vacancies) separately.  

\textcolor{black}{Following the previous studies of Refs. \citep{knolle19,kao21},
we also observe that a finite concentration of defects generically leads to a power-law divergence in the low-energy density of states (DOS).  Motivated by this robust result,  we then construct a comprehensive Griffiths-like scenario \citep{miranda05,vojta06,andrade09,vojta10,vojta14} and establish that (i) the DOS power-law exponent $\alpha$ is nonuniversal; (ii) $C/T$ and $\chi$ diverge with the same exponent $\alpha$; (iii) there is a nontrivial scaling for $C/T$ when $T/B \ll 1$; (iv) $1/T_{1}T$ follows a Korringa-like law.  All these findings are in accordance with the experimental results for \hlio.  Importantly,  this scenario does not  rely on the formation of random singlets \citep{bhatt82,zhou09,uematsu18,liu18,kimchi18a,kimchi18b,sanyal21}, which is a topologically trivial state,  unlike a disordered Kitaev spin liquid \citep{yamada21,  nasu20}.  For bond disorder,  however,  robust power-law singularities at low magnetic fields are linked to the destruction of the topological phase.  For vacancies,  we find that Kitaev's chiral spin-liquid phase survives up to a critical dilution, due to a nontrivial flux configuration,  before being replaced by a topologically trivial phase with power-law singularities.  This suggests that a half-quantized thermal Hall conductance might be detected experimentally in diluted Kitaev materials \citep{manni14,  do18,  do20,  baek20}.
}

\begin{figure}[t]
\begin{centering}
\includegraphics[width=1\columnwidth]{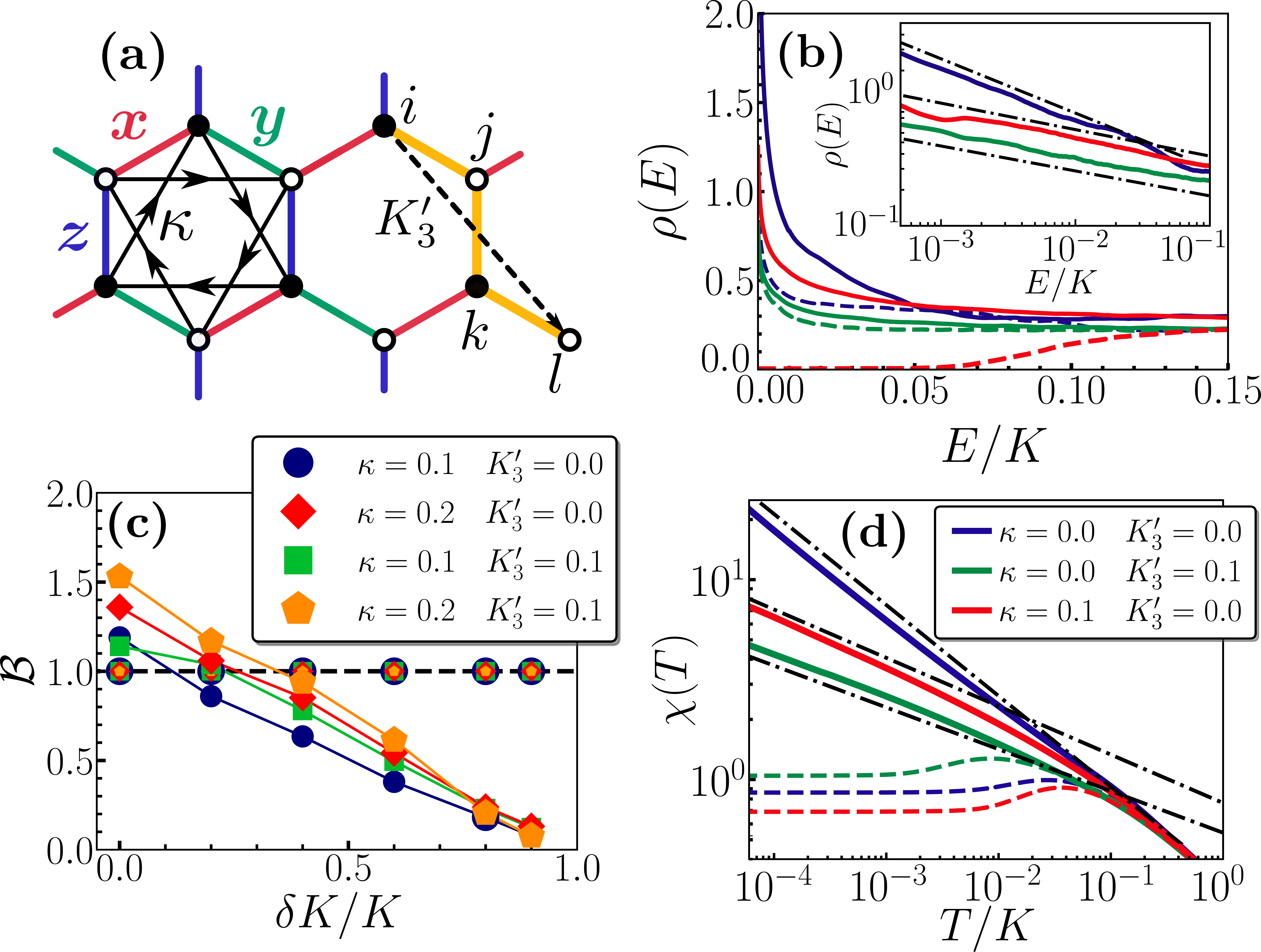}
\par\end{centering}
\caption{\label{fig:bond}Extended Kitaev model in Eq. \eqref{eq:kitaev_majoranas}
in the presence of bond disorder. (a) Links $x$, $y$, and $z$ in
the honeycomb lattice and the hopping between second $\left(\kappa\right)$
and third neighbors $\left(\kp\right)$. (b) DOS as
function of the energy for $\delta K=0.8$. Inset: log-log plot showing
the power-law divergence at low $E$. (c) Bott index as a function
of disorder. (d) Static uniform spin susceptibility as a function
of the temperature in a log-log plot for $\delta K=0.8$.  In (b)-(d)
full (dashed) curves correspond to random $(0)$ flux.  We consider the same parameters in (b) and (d).  \textcolor{black}{The dot dashed curves in (b) and (d) are power-law fits,  shifted with respect to the original curves,  with $\alpha=0.455(5),\,0.222(3),\,0.209(4)$}. We considered $L=30$ and $3 \times 10^{3}$ realizations of disorder.}
\end{figure}

\emph{Extended Kitaev model.}---As a minimal model to capture the
low-energy physics of \hlio, we consider an extended Kitaev model
on the honeycomb lattice \citep{zhang19}

\begin{equation}
\mathcal{H}=K\sum_{\left\langle ij\right\rangle _{\alpha}}\sigma_{i}^{\alpha}\sigma_{j}^{\alpha}+\kappa\sum_{\left\langle \left\langle ik\right\rangle \right\rangle }\sigma_{i}^{\alpha}\sigma_{j}^{\beta}\sigma_{k}^{\gamma}-\kp\sum_{\left\langle \left\langle \left\langle il\right\rangle \right\rangle \right\rangle }\sigma_{i}^{\alpha}\sigma_{j}^{\gamma}\sigma_{k}^{\gamma}\sigma_{l}^{\alpha},\label{eq:kitaev_spin}
\end{equation}
where $K$ is the usual Kitaev coupling and $\sigma_{i}^{\alpha}$
is the Pauli matrix at site $i$ with spin component $\left\{ \alpha,\beta,\gamma\right\} \in\left\{ x,y,z\right\} $.
$\left\langle ij\right\rangle _{\alpha}$ labels the nearest-neighbor
sites $i$ and $j$ along one of the three different links, see Fig.
\ref{fig:bond}(a). The three-spin term mimics the effects of an external
magnetic field and breaks time-reversal symmetry,  with $\kappa\propto h_{x}h_{y}h_{z}/\Delta_{2f}$, where $\Delta_{2f}$ is the two-flux gap (to be defined below)
\citep{kitaev06}. The four-spin interaction runs along a path of
length $3$, see Fig. \eqref{fig:bond}(a). While the $\kp$ term
can be generated perturbatively if one includes exchange couplings
beyond the pure Kitaev \citep{takikawa20}, we consider Eq. \eqref{eq:kitaev_spin}
as an effective low-energy theory \citep{song16}, and treat $\kp$
as one of the first allowed terms in the theory that preserves time-reversal
symmetry \citep{zhang19}. To restrict the model's parameter space,
we consider $K,\kappa,\kp>0$.

Equation \eqref{eq:kitaev_spin} is still amenable to the Kitaev's exact
solution \citep{kitaev06}.  We write the spin operator in terms
of four Majorana fermions $\sigma_{j}^{\alpha}=ib_{j}^{\alpha}c$,
and the Hamiltonian becomes
\begin{align}
\mathcal{H} & =-iK\sum_{\left\langle ij\right\rangle _{\alpha}}u_{ij}^{\alpha}c_{i}c_{j}-i\kappa\sum_{\left\langle \left\langle ik\right\rangle \right\rangle }u_{ij}^{\alpha}u_{kj}^{\beta}c_{i}c_{k},\nonumber \\
 & +i\kp\sum_{\left\langle \left\langle \left\langle il\right\rangle \right\rangle \right\rangle }u_{ij}^{\alpha}u_{kj}^{\beta}u_{kl}^{\alpha}c_{i}c_{i},\label{eq:kitaev_majoranas}
\end{align}
where $u_{ij}^{\alpha}=-u_{ji}^{\alpha}=ib_{i}^{\alpha}b_{j}^{\alpha}=\pm1$
is a conserved $\mathbb{Z}_{2}$ gauge field along the $\alpha$ bond
$\left\langle ij\right\rangle _{\alpha}$. The flux around each hexagonal
plaquette $p$ is a conserved quantity and may be written as $W_{p}=\prod_{\left\langle ij\right\rangle \in p}u_{ij}^{\alpha}$
\citep{kitaev06}. Since the fluxes commute with each other and with
the Hamiltonian in Eq. \eqref{eq:kitaev_majoranas}, once we fix the
link variable $u_{ij}^{\alpha}$ at each bond, thus defining a flux
sector, the problem can be solved exactly as a tight-binding model
of Majorana fermions and we obtain $\mathcal{H}=\sum_{\nu}\left(E_{\nu}-1/2\right)f_{\nu}^{\dagger}f_{\nu}.$
The operators $f_{\nu}$ are complex fermions operators (consisting
in the superposition of two Majorana operators \citep{suppl}) that
label the eigenstate with energy $E_{\nu}$ in a given flux sector.

For the pure Kitaev model, $\kappa=\kp=0$, in the absence of disorder,
we have a $0$-flux ground state, i.e., $W_{p}=+1$ for all plaquettes.
The two-flux gap is the difference between the ground state energies
of a system with a single flipped link variable and the original reference state because a single bond flip creates a pair of fluxes in neighboring hexagons \citep{kitaev06,knolle14}.
The ground state flux state depends on $\kp$ \citep{zhang19}. For our parameter regime,  we find that for $\kp\gtrsim K/8$ the ground state comprises
one flux per plaquette,  i.e., $W_{p}=-1$ for all plaquettes \citep{suppl}, but we do not explore this transition in this work.  We consider
finite clusters of linear size $L$, with periodic boundary conditions.
Because we have two sites per unit cell, the total number of sites
is $N=2L^{2}$.


\emph{Bond disorder and random flux}.---We now add disorder to the
model in Eq. \eqref{eq:kitaev_majoranas}. Specifically, we consider
a binary bond disorder for \textcolor{black}{all} Kitaev couplings setting $K\to K\pm\delta K$, with probability $0.5$ to generate either a weak bond, $\left(K-\delta K\right)$, 
or a strong bond $\left(K+\delta K\right)$.  For simplicity,  we assume the couplings $\kappa$ and $\kp$ to be homogenous.  In terms of Majorana fermions, this problem translates into a random hopping problem in a bipartite lattice.  For $\kappa=\kp=0$,
it is rigorously known that the DOS, $\rho\left(E\right)=\sum_{\nu}\delta\left(E-E_{\nu}\right)/N$,
has a low-$E$ divergence $\rho\text{\ensuremath{\left(E\right)}}\sim\exp\left(-c\left|\ln E\right|^{2/3}\right)/E$,
where $c$ is a positive constant \citep{motrunich02,  sanyal16}. Nevertheless,
this divergence occurs only at asymptotically low-energy scales, eluding
even large-scale numerical simulations \citep{motrunich02,  sanyal16}. This
fact probably places it outside the experimentally accessible regimes
for magnetic materials.

After fixing the flux state,  we numerically diagonalize Eq. \eqref{eq:kitaev_majoranas}.
We find that the static flux configuration is sensitive to strong disorder \citep{zschocke15}.  Specifically,  we observe that for $\delta K\gtrsim0.6$
the ground state energies of the different flux sectors become comparable within the error
bars \citep{suppl}. This implies that $\Delta_{2f}\to0$ as $\delta K\to1$,
as in recent quantum Monte Carlo results \citep{nasu20,nasu21}.  Therefore,  we consider the random flux background a competitive variational flux state.  In this state,  we randomly set $u^{\alpha}_{ij} = \pm 1$ at each link with equal probability.

Figure \ref{fig:bond}(b) shows the averaged DOS for $\delta K=0.8$.
For $\kp=\kappa=0$, both the $0$-flux and random-flux states produce
a diverging power-law behavior at low $E$: $\rho\left(E\right)\sim E^{-\alpha}$,
in accordance with the results of Refs. \citep{knolle19,kao21}.  If
one assumes that the flux degrees of freedom are frozen,  it follows that
$C/T\sim T^{-\alpha}$ \citep{knolle19,kao21}. The power-law
exponent $\alpha$ is nonuniversal and it depends continuously on
the model parameters; see Fig. \ref{fig:bond}(b) \citep{suppl}.
The evolution of $\alpha$ with $\kappa$ is particularly sensitive to the flux
sector. While the random-flux sector experiences a reduction
of $\alpha$ as $\kappa$ increases,  the $0$-flux state
displays a gap in the Majorana spectrum.  This gap is reminiscent of the topological gap present in the clean case \citep{kitaev06}. 

To further explore the effects of bond disorder on the thermodynamic
response, we also calculate the static uniform susceptibility employing
the adiabatic approximation \citep{baskaran07,  knolle14,  knolle15,suppl}. We show
sample results for $\kappa=\kp=0$ and $\delta K=0.8$ in Fig. \ref{fig:bond}(d).  The random-flux sector shows a diverging susceptibility $\chi\sim T^{-\alpha}$
at low $T$,  with $\alpha$ the DOS exponent,  in line with the results
for \hlio ~\cite{kitagawa18}. The $0$-flux state, on the other
hand, shows a finite $\chi$ at low $T$.  We can trace back these behaviors to the value of $\Delta_{2f}$.  In the $0$-flux state,  $\Delta_{2f}$ remains
finite,  albeit smaller,  for $\delta K>0$ \citep{nasu21,yamada20},
and the susceptibility goes to a constant for $T < \Delta_{2f}$.
In the random-flux state,  $\Delta_{2f}\to0$ and $\chi\left(T\right)$
follows $\rho\left(E\right)$ at low energies. 

These two flux states also manifest differently in the topological
properties of the system; see Fig. \ref{fig:bond}(c).  To capture a nontrivial topological phase,  we compute the Bott index,  which is equivalent to the Chern number in periodic systems. Still, it is more conveniently implemented in systems lacking translational invariance \citep{hastings11,agarwala17,huang18,suppl}.
The $0$-flux state shows a stable topological phase up to $\delta K\to1$
due to the finite topological gap in the Majorana spectrum present
in $\rho\left(E\right)$ \citep{yamada20,nasu20}. This is
similar to what is observed in disordered two-dimensional disordered
Chern insulators \citep{onoda07,prodan10,garcia14,castro15}.   However, there is a pileup of low-energy states in the random-flux state, even for $\delta K =0$, and the topological phase is destroyed for all $\delta K$. We complement the Bott index results with an investigation of the level spacing statistics \citep{prodan10,oganesyan07,atas13,suppl},  and the results are entirely consistent.

Based on our results,  we construct the following scenario for disordered
Kitaev materials.  Taking $\kappa$ to mimic the effects of an
external magnetic field,  the experimental results observed in \hlio
~\cite{kitagawa18} can be described by Eq. \eqref{eq:kitaev_majoranas},
augmented by bond disorder, only if one assumes a random-flux state
\citep{knolle19,kao21}. This, in turn, implies that power-law singularities
at zero fields are associated with a topologically trivial phase in a finite field also displaying power-law singularities but with a smaller exponent.


\emph{Griffiths-like response.}---We now present a physical mechanism behind the power-law singularities in the DOS.  Although
this is a crossover regime \citep{motrunich02,  sanyal16},  the fact
that it emerges for distinct choices of disorder distributions \citep{knolle19,kao21}
suggests a more general picture. 

Power-law distribution of energy scales is commonly observed in
the vicinity of quantum critical points in disordered systems \citep{vojta06,miranda05,andrade09,vojta10,vojta14},
in the so-called Griffiths phase. We exploit this similarity
and propose the following mechanism.  Suppose a rare region (droplet) of linear size $\ell$ contains only weak bonds at its boundaries. The probability of finding such cluster inside the bulk is $P\left(\ell\right)\propto\exp\text{\ensuremath{\left[b\ln\left(p\right)\ell\right]}}$,
where $p$ is the probability of finding a single weak bond and $b>0$
is a constant. For a completely disconnected region, $\delta K \to K$,
a finite-size gap appears in the Majorana spectrum $\Delta\left(\ell\right)\propto\exp\left[-a\ell\right]$,
with $a>0$ another constant. This gap comes from the hybridization
of the localized states at the edges of this cluster \citep{suppl}.
The contribution to the density of states coming from these rare regions
is $\rho\left(E\right)=\int d\ell\,P\left(\ell\right)\delta\left[E-\Delta\left(\ell\right)\right]\sim E^{-\alpha}$,
with $\alpha=1+\left(b/a\right)\ln\left(p\right)$. Therefore, weakly
coupled clusters give rise to a power-law singularity in the DOS.
For even lower temperatures,  we eventually flow away from this crossover regime toward the asymptotic result $\rho\text{\ensuremath{\left(E\right)}}\sim\exp\left(-c\left|\ln E\right|^{2/3}\right)/E$ \citep{motrunich02,  sanyal16}.

We extend this Griffiths phase analogy and calculate
the leading low-$T$ contribution to several physical observables in the limit
of \textcolor{black} {frozen flux configurations},  such that we can link the spin excitations
solely to $\rho\left(E\right)$. For instance, we can estimate the
number of free clusters as $n\left(T\right)\sim\int_{0}^{T}\rho\left(E\right)dE\sim T^{-\alpha+1}$.
This leads to a finite low-$T$ entropy for the spins $S\sim n\left(T\right)\ln2$
and thus $C/T\sim T^{-\alpha}$. Analogously, the uniform spin susceptibility
can be estimated as $\chi\left(T\right)\sim n\left(T\right)/T\sim T^{-\alpha}$, 
which eventually overcomes any regular contribution from the bulk.  \textcolor{black}{Importantly,  this result does not rely on the adiabatic approximation}.
The imaginary part of the dynamical susceptibility is given by $\chi^{\prime\prime}\left(\omega\right)\sim\int\delta\left(\omega-E\right)\rho\left(E\right)dE\sim\omega^{-\alpha}$.
Because the cluster excitations are essentially local, we may write
the NMR spin-relaxation rate $1/T_{1}$ as \citep{coleman15} $1/T_{1}T\sim\chi^{\prime\prime}\left(\omega_{o}\right)/\omega_{o}\sim\omega_{o}^{-\alpha-1}$,
where $\omega_{o}$ is the nuclear resonance frequency. Therefore,
$1/T_{1}T$ remains finite down to very low temperatures. Lastly, we
can also discuss the curious data scaling encountered in Ref. \citep{kitagawa18}:
$C/T\sim B^{-3/2}T$ for $T<B$, where $B$ is the magnetic field.
First, we write \citep{kitaev06} $\kappa\sim B^{3}/\Delta_{2f}^{2}\sim B^{3}/T^{2}$,
setting $T$ as the low-energy scale in this regime. Because $T\ll\kappa$,
we employ a Sommerfeld-like expansion and write $C\sim\rho\left(\kappa\right)T\sim\kappa^{-\alpha}T\sim B^{-3\alpha}T^{1+2\alpha}$.
For $\alpha=1/2$, we obtain the experimentally observed scaling. 

Therefore, a disordered extension of Kitaev's spin liquid provides
a consistent scenario to the experimental results observed in \hlio \: 
once we combine a power-law low-energy DOS with standard Griffiths-like
arguments. However, such a scenario is incompatible with a topological
nontrivial phase for bond disorder alone because it requires a
random-flux state. 

\begin{figure}[t]
\begin{centering}
\includegraphics[width=1\columnwidth]{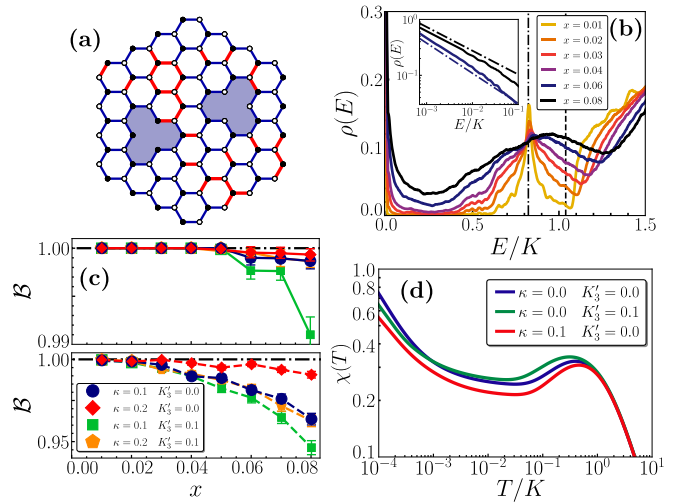}
\par\end{centering}
\caption{\label{fig:vac}Diluted extended Kitaev model,  Eq. \ref{eq:kitaev_majoranas}.  (a) Bound-state flux configuration. The shaded $l=12$ plaquettes show the binding of a flux by each vacancy. (b) DOS as function of the energy for $\kappa=0.1K$ and $\kp=0$  Inset:
log-log plot showing the power-law divergence at low $E$ for $x\ge 0.06$.  \textcolor{black}{The dot-dashed curves are power-law fits,  shifted with respect to the original curves,  with $\alpha=0.496(1),\,0.445(3)$}. (c) Bott index as a function of the dilution for the bound flux ($0$ flux) in the upper (lower) panel. (d) Log-log plot of $\chi \times T$ in the bound-flux sector for $x=0.04$.  We use the same parameters as in Fig.  \ref{fig:bond}}
\end{figure}

\emph{Site dilution and unpaired spins.}---We now introduce vacancies
in the extended Kitaev model, Eq. \eqref{eq:kitaev_majoranas}. Specifically,
we remove a fraction $x$ of spins from the system. To avoid trivial
zero-energy modes, we remove exactly $xN/2$ spins from each sublattice.
In the limit $x\ll1$, a vacancy binds a flux to it \citep{willans10}:
as one loops the impurity plaquette, $W_{p}=-1$;  see Fig.
\ref{fig:vac}(a).  Because we work at finite $x$,  we consider both $0$-flux and bound-flux
states \citep{kao21}.  In the $0$ (bound) flux,  $W_{p}=+1(-1)$ in the vacancy plaquette.  For all other honeycomb loops,  $W_{p}=+1$; see \ref{fig:vac}(a).  We find that the random-flux state is not
a competitive ground state for $x\lesssim0.1$ \citep{suppl}. 

Unlike the bond disordered case,  a small dilution is sufficient to induce a pileup of low-energy states, similar to what is observed in graphene \citep{pereira06,pereira08,hafner14},  independently of the flux sector \citep{kao21,suppl}.  A clearer distinction between the different \textcolor{black}{flux configurations} emerges for $\kappa \neq 0$.  In Fig. \ref{fig:vac}(b) we show the DOS for the bound-flux state,  $\kappa=0.1K$,  $\kp=0$,  and several values of dilution $x$.  For $x \lesssim 0.05$, we observe a localized level inside the clean gap. The larger the $\kappa$,  the more well-defined this state is.  As we increase $x$, we reduce the impurity distance---its typical value scales as $1/\sqrt{x}$---enhancing the overlap between the impurity states, which gives rise to an impurity band inside the clean topological gap,  similar to what is observed in disordered Chern insulators \citep{onoda07,  prodan10,  castro15}.  The resulting phase is a topologically trivial phase with power-law singularities. See the inset of Fig. \ref{fig:vac}(b).

To see the effects of the in-gap states on the topological properties of the system,  we compute the Bott index; see Fig. \ref{fig:vac}(c).  In the small $\kappa$ regime,  $\mathcal{B}$ is no longer quantized for the $0$-flux \textcolor{black}{state} if $x > 0.02$.  For the bound-flux state,  however,  $\mathcal{B}$ remains pinned to an integer up to $x \approx 0.05$ (this critical value depends on $\kappa$ \cite{suppl}).  In both cases,  the clean topological gap is the same, and this extra robustness of the bound-flux state is rooted in the in-gap state at finite energy shown in Fig. \ref{fig:vac}(b).  The existence of this state can be understood as follows.  Consider the impurity plaquette as a $l=12$ tight-binding chain with nearest-neighbor hopping only (the vacancy plaquette has a length $l=12$ rather than $l=6$ for the elementary honeycomb one).  The spectrum of this problem has (does not have) a gap if the chain binds (does not bind) a flux.  For finite $x$,  the impurity states go into this level,  ensuring the localization of the impurity states around the vacancy for small $x$.  For larger concentrations,  other impurities configurations become relevant,  e.g., a pair of neighboring vacancies \citep{suppl},  and more states inside the topological gap are populated.  This suggests that a topological phase could be stable in the diluted system for an external field that is large enough to quantize $\mathcal{B}$ for a given $x$,  but not
too large as to move the system away from the bound-flux state \citep{willans10,kao21}.  The presence of this topological phase might be probed experimentally using the thermal Hall conductance \citep{kasahara18b,yokoi21,czajka21}. Despite being challenging, these measurements could be relevant both to H$_{3}$LiIr$_{2}$O$_{6}$ \citep{kitagawa18} and Ir-doped RuCl$_{3}$ \citep{do18,  do20,  baek20}.

In Fig. \ref{fig:vac}(d) we show sample results for the static uniform
spin susceptibility $\chi\left(T\right)$. We observe a mild increase
in $\chi\left(T\right)$ for the bound flux,  with similar results for the $0$ flux.  A bona fide power-law
divergence is present only at much lower temperatures \citep{suppl}. This behavior is due to the existence of unpaired bonds \citep{nasu21}. By removing a site,  we automatically leave its three nearest neighbors disconnected along one bond.  Such unpaired bonds automatically display
$\Delta_{2f}=0$, and they dominate the low-$T$ behavior of $\chi\left(T\right)$. However, since the fraction of unpaired spins is equal to $x$, at least for small $x$, their overall contribution
is masked by the remaining $1-x$ fraction of bulk spins that give a finite contribution to $\chi\left(T\right)$ if $T<\Delta_{2f}$ \citep{suppl}. This is also in line with the Knight shift measurements reported in \citep{kitagawa18}: spins far away from the defects produce a regular flat contribution to the local spin susceptibility, whereas spins around a vacancy give a singular response.  Because the condition $\Delta_{2f}=0$ is automatically satisfied by these unpaired spins, the Griffiths-like arguments discussed previously apply directly here, regardless of the considered static flux background. As a closing remark, we stress that the asymptotic results for $\chi\left(T\right)$ calculated in Ref. \citep{willans10} are only relevant for large fields, where the magnetic length is smaller than the typical interimpurity distance, and the single vacancy limit holds.

\emph{Conclusions.}---We investigated an extended Kitaev model, Eq.
\eqref{eq:kitaev_majoranas}, in the presence of defects.  We find the emergence of a singular power-law density
of states at low energies \citep{knolle19,kao21}  with a nonuniversal exponent.
We then construct a phenomenological scenario for our numerical findings by discussing this power-law distribution of energy scales in terms of a Griffiths-like phase.  Our results provide a consistent scenario to the experimental observations for \hlio~\citep{kitagawa18}, \textcolor{black}{and we expect them to describe other diluted Kitaev materials as RuCl$_{3}$ \citep{do18,  do20,  baek20} and the Iridates \citep{manni14,hk_depl14}}. 

From a theoretical perspective,  this unanticipated link deserves further study since a Griffiths-like phenomenology appears naturally in a random-singlet phase \citep{bhatt82,zhou09,uematsu18,liu18,kimchi18a,kimchi18b,sanyal21}.
In the absence of disorder, a valence-bond crystal and the Kitaev
spin liquid are separated by a quantum phase transition \citep{vojta18,majumder18}.
Our work points toward an exciting evolution of this critical point with the disorder.

In the presence of a time-reversal breaking term, we find that the topological properties of the system are sensitive both to the static flux background and to the particular choice of disorder. For bond disorder, the power-law singularities are robust only if one assumes a random-flux background, implying a lack of a topological spin-liquid phase.  However, the power-law singularities survive at weak external magnetic fields for small concentrations of vacancies.  They are eventually quenched at larger fields,  where a topological phase with chiral Majorana edge modes emerges. The stability of this topological phase comes from the fact that a vacancy binds a flux to it, which helps protect the clean topological gap in the Majorana spectrum. Our results indicate that diluted Kitaev materials are promising candidates to display Kitaev's chiral spin-liquid phase in weak to moderate magnetic fields.  

\begin{acknowledgments}
We acknowledge useful discussions with Pedro C{\^o}nsoli,  Jos{\'e} Hoyos, Lukas Janssen, Eduardo Miranda,  Jo{\~a}o Augusto Sobral,  and Matthias Vojta.  V.D.
acknowledges the support from the University of S\~ao Paulo. E.C.A.
was supported by CNPq (Brazil), Grants No. 406399/2018-2 and 302994/2019-0,
and FAPESP (Brazil), Grant No. 2019/17026-9. 
\end{acknowledgments}



%

\end{document}


\title{Supplementary information for:\\
 ``Disorder, Low-Energy Excitations, and Topology in the Kitaev Spin Liquid'' }
\author{Vitor Dantas}
\affiliation{Instituto de F\'isica de S\~ao Carlos, Universidade de S\~ao Paulo,
C.P. 369, S\~ao Carlos, SP, 13560-970, Brazil}
\author{Eric C. Andrade}
\affiliation{Instituto de F\'isica de S\~ao Carlos, Universidade de S\~ao Paulo,
C.P. 369, S\~ao Carlos, SP, 13560-970, Brazil}
\date{\today}

\maketitle
\section{Diagonalization of extended Kitaev model and thermodynamics}

We start from Eq.  (2) in the main text with the extended Kitaev model written in terms of Majorana fermions

\begin{align}
\mathcal{H} & =-iK\sum_{\left\langle ij\right\rangle _{\alpha}}u_{ij}^{\alpha}c_{i}c_{j}-i\kappa\sum_{\left\langle \left\langle ik\right\rangle \right\rangle }u_{ij}^{\alpha}u_{kj}^{\beta}c_{i}c_{k},\nonumber \\
 & +i\kp\sum_{\left\langle \left\langle \left\langle il\right\rangle \right\rangle \right\rangle }u_{ij}^{\alpha}u_{kj}^{\beta}u_{kl}^{\alpha}c_{i}c_{l}.\label{eq:kitaev_majoranas}
\end{align}
The nearest-neighbor hopping $K$ is the usual Kitaev coupling, and $\left\langle ij\right\rangle _{\alpha}$ labels the nearest-neighbor
sites $i$ and $j$ along one of the three different links of the honeycomb lattice.  The next-nearest neighbor hopping $\kappa$ mimics the effects of an external
magnetic field -- $\kappa\propto h_{x}h_{y}h_{z}/\Delta_{2f}$, where
$\Delta_{2f}$ is the two-flux gap -- and breaks time-reversal symmetry
\citep{kitaev06}. The third neighbor hopping $\kp$ runs along a path of
length $3$ as shown in Fig.  1(a)  of the main text \citep{zhang19}.  $u_{ij}^{\alpha}=-u_{ji}^{\alpha}=\pm1$
is a conserved $\mathbb{Z}_{2}$ gauge field along the bond $\left\langle ij\right\rangle _{\alpha}$. The flux around each hexagonal plaquette $p$ is a conserved quantity and may be written as $W_{p}=\prod_{\left\langle ij\right\rangle \in p}u_{ij}^{\alpha}$
\citep{kitaev06}. Since the fluxes commute with each other and with
the Hamiltonian in Eq. \eqref{eq:kitaev_majoranas}, once we fix the
link variables $u_{ij}^{\alpha}$ at each bond,  thus defining a flux
configuration,  the problem can be solved exactly as a tight-binding model
of Majorana fermions.

To diagonalize Eq. \eqref{eq:kitaev_majoranas},  we rewrite it in the following matrix form
\begin{align}
\mathcal{H}& = \frac{i}{2}\label{Hmaj general}
\begin{pmatrix}
c_A & c_B
\end{pmatrix}
\begin{pmatrix}
F & M\\
-M^T& -D
\end{pmatrix}
\begin{pmatrix}
c_A\\c_B
\end{pmatrix}, 
\end{align}
with the matrix $M_{ij} = K u_{ij} - K_3^\prime u_{il}^\alpha u_{kl}^\beta u_{kj}^\alpha$ defining the hopping between different sublattices. The hopping in the same sublattice is represented by the matrix elements $F_{ik} = \kappa u_{ij}^\alpha u_{kj}^\beta$ and $D_{ik} = \kappa u_{ij}^\alpha u_{kj}^\beta$. Note that we generically have $F\neq D$ due to the sublattice symmetry breaking induced by a generic flux configuration. Now we introduce the complex fermion operators $d$ and $d^\dagger$, which are related to the Majorana fermions by $d = \pa{ c_A + ic_B}/2$ and $d^\dagger = \pa{c_A - ic_B}/2$. In this complex fermion basis, the Hamiltonian assumes the Bogoliubov de-Gennes form
\begin{align}\label{Hcomplex}
\mathcal{H}  = \frac{1}{2}\;
\begin{pmatrix}
d^\dagger & d
\end{pmatrix}
\begin{pmatrix}
h&\Delta\\
\Delta^\dagger&-h^T
\end{pmatrix}
\begin{pmatrix}
d\\
d^\dagger
\end{pmatrix},
\end{align}
with $\Delta$ and $h$ defined in terms of $M$, $D$, and $F$ as
\begin{align}
&\Delta = (M^T - M) + i(F+D),\\
&h = (M + M^T) + i(F - D).
\end{align}

The Bogoliubov quasiparticle operators $f$ and $f^\dagger$ are related to $d$ and $d^\dagger$ via 
\begin{align}\label{bogtrans}
\begin{dcases}
&d_\nu = \sum_{\lambda} X^{T}_{\nu\lambda} f_\lambda^{} + Y_{\nu\lambda}^\dagger f_\lambda^\dagger, \\
&d_\nu^\dagger = \sum_{\lambda} Y^{T}_{\nu\lambda} f_\lambda^{} + X_{\nu\lambda}^\dagger f_\lambda^\dagger.
\end{dcases}
\end{align}
$X_{\nu\lambda}$ and $Y_{\nu\lambda}$ are the Bogoliubov matrices,  corresponding to the occupied and empty states, respectively.  In terms of the Bogoliubov quasiparticles,  Eq.  \eqref{eq:kitaev_majoranas} becomes diagonal
\begin{align}
\mathcal{H} = \sum_\nu\pa{f_\nu^\dagger f_\nu - \frac{1}{2}} E_\nu. 
\end{align}

The density of states (DOS) can be readily calculated as $\rho\left(E\right)=\sum{_\nu}\delta\left(E-E_{\nu}\right)/N$.  With the DOS,  we can calculate thermodynamic quantities,  as the specific heat.   In the presence of disorder, we observe $\rho\left(E\right) \sim E^{-\alpha}$ \citep{knolle19,  kao21},  with a  non-universal exponent $\alpha$,  unlike the graphene-like spectrum of the clean system.  If we assume that at low temperatures the fluxes degrees of freedom are frozen,  the specific heat is given by $C\left(T\right) \sim \partial \left[\int dE E^{1-\alpha} f\left(E\right)\right]/\partial T \sim T^{1-\alpha}$,  in accordance with our Griffiths-like scenario.  Here, where $f(E)$ is the Fermi-Dirac distribution with zero chemical potential: $f(E) = 1/(e^{E/T}+1)$. 

The calculation of the static spin susceptibility is more involved.  Using the Majorana representation of Eq.  \eqref{eq:kitaev_majoranas},  a spin-flip causes the flip of a bond,  creating a pair of fluxes with energy $\Delta_{2f}$ on top of the Majorana excitations.  We now describe the approximation we employ to calculate $\chi\left(T\right)$.


\subsection{Adiabatic approximation and the static susceptibility}

The dynamical spin structure factor is defined by:
\begin{align}\label{Struc_1}
S(\textbf{q},\omega) = \frac{1}{N}\sum_{ij}\sum_{\alpha\beta}e^{-i\textbf{q}\cdot(\textbf{r}_i - \textbf{r}_j)}S_{ij}^{\alpha \beta}(\omega),
\end{align}
where $S_{ij}^{\alpha\beta}$ is the Fourier transform of the spin-spin correlation function:
\begin{align}
S_{ij}^{\alpha \beta}(\omega) = \int_{-\infty}^{\infty}dt\;e^{i\omega t} \mean{\sigma^\alpha_i(t)\sigma^\beta_j(0)}.
\end{align}
The spin-spin correlation is ultra short-ranged,  with only on-site and nearest-neighbors correlations non-zero \citep{baskaran07}. Due to the $C_3$ symmetry of the problem,  we only calculate its $zz$ component (in the disordered case, this symmetry holds after disorder averaging).

The calculation of $S_{ij}^{\alpha \beta}(\omega)$ can be performed exactly for the clean system \citep{knolle14, knolle15}. However, the numerical calculation in this scheme is quite challenging because it requires the overlap between the original flux configuration and the new one with a bond flipped.  To reduce the computational cost, we work within the adiabatic approximation \citep{knolle19,  knolle15, zschocke15}.  In this approach,  we assume this overlap to be always finite,  and we perform all calculations in the new flux configuration with a single bond flipped at $\textbf{q} = 0$:
 \begin{align}\label{struc_ad}
 S^{zz}(\textbf{q} = 0, \omega)& = \sum_\nu \delta(\omega - E_\nu - \Delta_{2f})\abs{X_{\nu o}^{} }^2 f(-E_\nu) \nonumber \\ 
 & +  \sum_\nu \delta(\omega + E_\nu - \Delta_{2f})\abs{Y_{\nu o}^{} }^2 f(E_\nu) ,
 \end{align}
The index $o$ specifies the flipped bond position.  We average over different positions $o$ for a given disorder and flux configuration in the disordered case.  The static spin susceptibility is finally given by \cite{coleman15}
\begin{align}\label{chi_uni_final}
\chi(T) = &\int_{-\infty}^{\infty} d\omega \; S(\textbf{q} = 0,\omega) \;\frac{1-e^{-\omega/T}}{\omega}, \nonumber \\
 = &\sum_{\nu}  \abs{X_{\nu o}^{} }^2 f(-E_\nu) \;\frac{1-e^{-(E_\nu + \Delta_{2f})/T}}{E_\nu + \Delta_{2f}}  \nonumber \\
&+ \abs{Y_{\nu o}^{} }^2 f(E_\nu) \;\frac{1-e^{(E_\nu - \Delta_{2f})/T}}{-E_\nu + \Delta_{2f}}.
\end{align}
To calculate $\chi(T)$, we need not only to diagonalize Eq.  \eqref{eq:kitaev_majoranas} with a bond flipped but also we need to know the value of the two flux gap $\Delta_{2f}$: the energy difference between the configuration with a single link variable flipped with respect to the reference flux state.

\begin{figure}[t]
	\centering
	\includegraphics[width=1.0\linewidth]{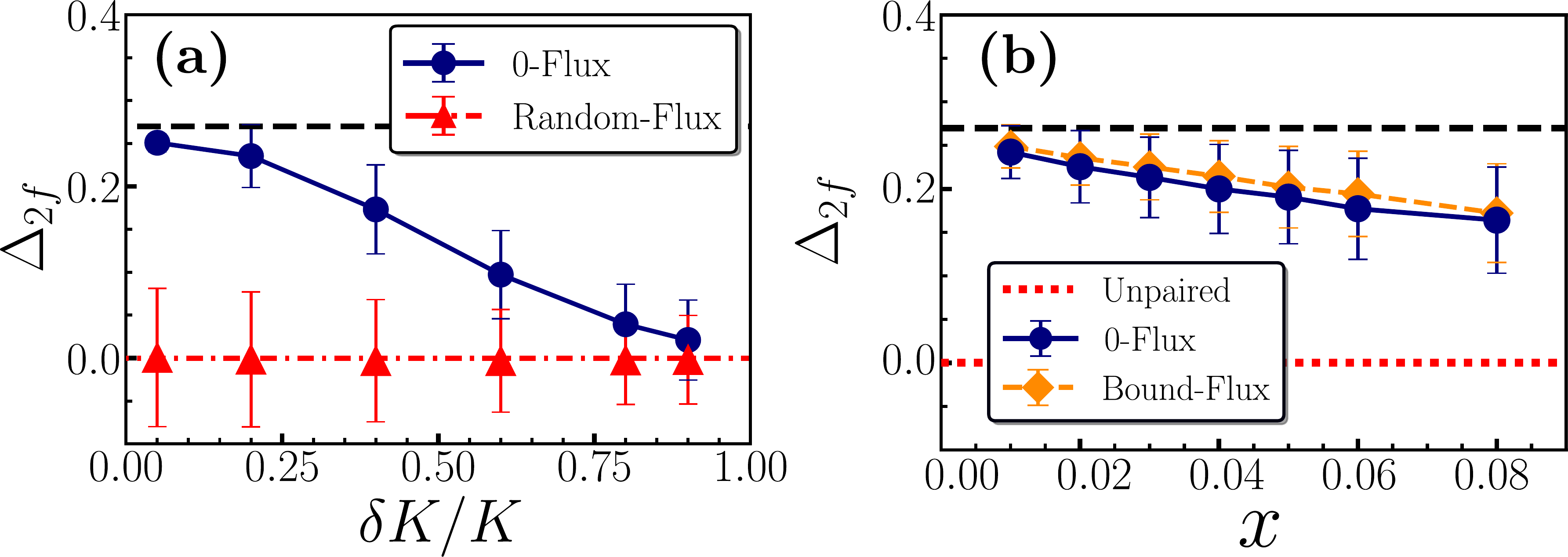}
	\caption{The flux pair gap $\Delta_{2f}$ as a function of disorder for $\kappa=\kp=0$.  The dashed line corresponds to its clean value,  $\Delta_{2f} \approx 0.27K$. (a) Bond disorder with $\Delta_{2f}$ as a function of the strength of the bond disorder $\delta K/K$. (b) Site dilution with $\Delta_{2f}$ as a function of the vacancy concentration $x$.  $\Delta_{2f} = 0$ for the unpaired spins.  We considered $L=30$ and $3\times10^3$ realizations of disorder.}
	\label{fig:figs1}
\end{figure}

For the bond disorder case,  Fig. \ref{fig:figs1}(a), the average value of $\Delta_{2f}$ vanishes in the random-flux state for all values of $\delta K/K$ and we set $\Delta_{2f}=0$ in our calculations.   For the $0$-flux state,  $\Delta_{2f}$ is finite and approaches zero in the limit of strong disorder,  $\delta K/K \to 1$,  \citep{yamada20,  nasu21}.  For the site dilution case, there are two contributions to $\Delta_{2f}$ coming from the paired and unpaired spins \citep{nasu21}.  An unpaired spin is located in a site that misses one bond due to the dilution of the nearest neighbor.  For the paired spins,  we have $\Delta_{2f}$ finite and slowly diminishing with the dilution $x$,  Fig. \ref{fig:figs1}(b).  For the unpaired spins,  $\Delta_{2f} = 0$ by definition since there is no bond to flip.  For $x \le 0.1$,  we observe a fraction of $x \,(1-x)$ unpaired (paired) spins.  Our system also has orphan spins,  spins missing all their nearest neighbors.  In the current model, they are of order $\mathcal{O}\left(10^{-3}x\right)$,  which implies that their contribution to the physical observables is negligible in the experimentally relevant temperature range.  As shown in the main text and the following sections,  $\chi(T)$ obtained within the adiabatic approximation agrees well with the expected results based on our Griffiths-like arguments,  suggesting this is a good approach for the problem at hand.

\subsection{Topology: Bott index and level spacing}

Without time-reversal symmetry,  $\kappa \neq 0$,  the extended Kitaev model is topologically non-trivial in the clean limit \citep{kitaev06,  zhang19,  takikawa20}.  To characterize the distinct topological phases,  we calculate the Bott Index, which is equivalent to the Chern number in periodic systems. Still, it is more conveniently implemented in systems lacking translational invariance \citep{hastings11,agarwala17,huang18}.  The Bott Index is defined as:
\begin{align}
\mathcal{B} = \frac{1}{2\pi}\Im \left\{\Tr\co{\log \pa{ VUV^\dagger U^\dagger }}  \right\},
\end{align}
where the matrices $U$ and $V$ are given by
\begin{align}
&P e^{2\pi i R_x/L} P = \begin{pmatrix}
X^* & Y^* \\
Y & X 
\end{pmatrix}
\begin{pmatrix}
0&0\\ 
0&U
\end{pmatrix} 
\begin{pmatrix}
X^T &Y ^\dagger \\
Y^T &X^\dagger 
\end{pmatrix},  \\
&P e^{2\pi i R_y/L} P = 
\begin{pmatrix}
X^*&Y^* \\
Y&X 
\end{pmatrix}
 \begin{pmatrix}
0&0\\ 
0&V
\end{pmatrix} \begin{pmatrix}
X^T &Y ^\dagger \\
Y^T &X^\dagger 
\end{pmatrix}. 
\end{align}
Here,  $P$ is the projector onto the occupied states:
\begin{align}
P = 
\begin{pmatrix}
X^* & Y^* \\
Y & X 
\end{pmatrix}
\begin{pmatrix}
0&0\\ 
0&\mathds{1}
\end{pmatrix} 
\begin{pmatrix}
X^T &Y ^\dagger \\
Y^T &X^\dagger 
\end{pmatrix}, 
\end{align}
with $X$ and $Y$ defined in \eqref{bogtrans}.  The operators $R_x$ and $R_y$ are diagonal matrices whose entries correspond to the unit cell positions $x_i$ and $y_i$, respectively. To increase the stability of the numerical algorithm, we perform a singular value decomposition on $U$ and $V$: $U = \Sigma S \Theta^\dagger$, where $S$ is diagonal, and $\Sigma$ and $\Theta$ are unitary matrices.  With this,  we redefine the projected matrices as $\widetilde{U}\equiv \Sigma\Theta^\dagger$, which is equivalent to a scaling transformation and does not change the Bott Index \cite{huang18}.

For the disordered system we calculate the Bott Index averaged over $N_s$ samples,  $\mathcal{B} = \sum_{s = 1}^{N_s} \mathcal{B}_{s}/{N_s}$.  Our numerical algorithm only returns values of $\mathcal{B}_s$ that are integers,  reinforcing the robustness of the method.  Therefore,  the averaged value of the Bott index corresponds to the proportion of topologically non-trivial samples found for a particular set of parameters.  

\begin{figure}[t]
	\centering
	\includegraphics[width=0.9\linewidth]{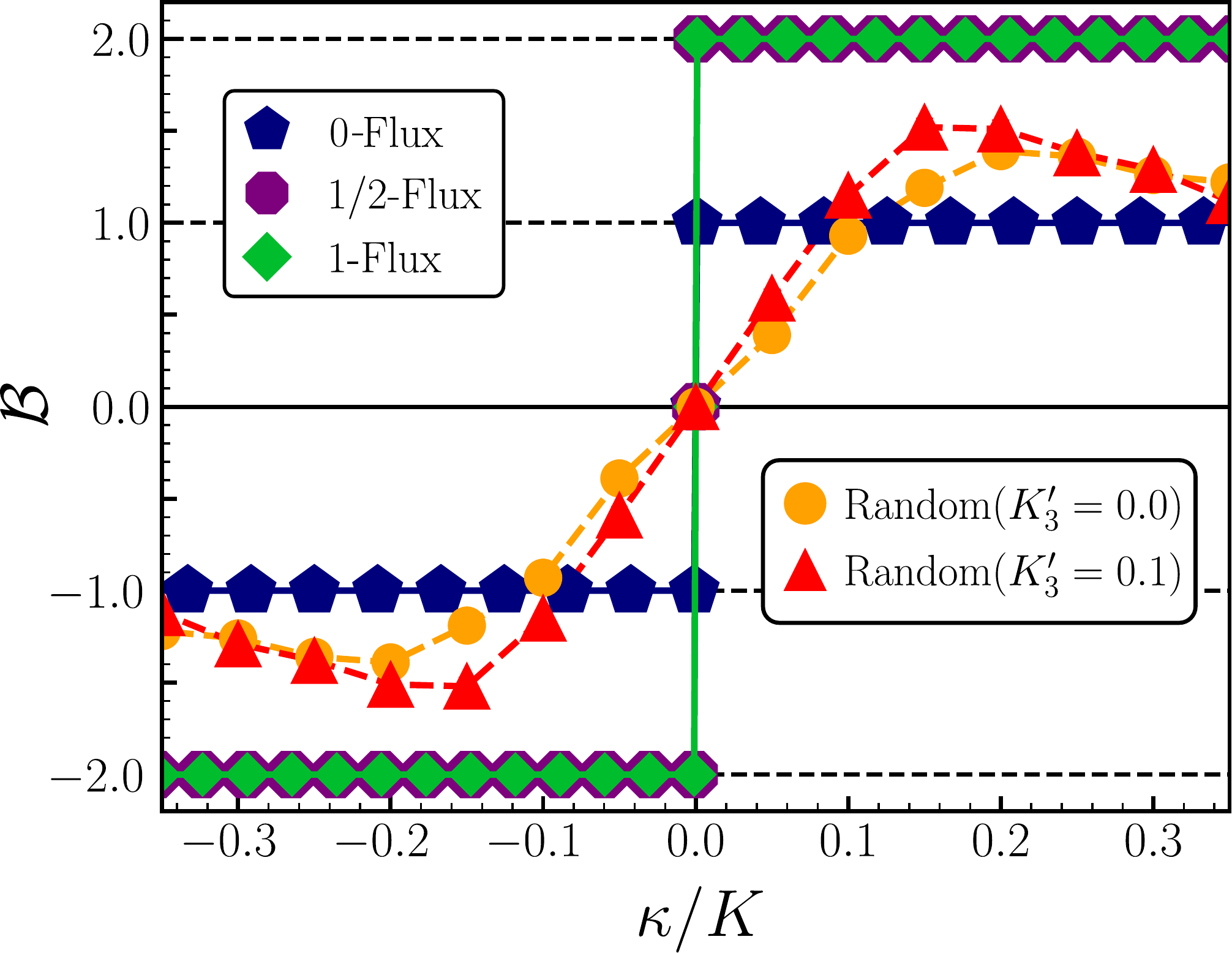}
	\caption{Bott Index $\mathcal{B}$ as a function of $\kappa$ for the clean case,  $\delta K=0$.  We show results for different flux configurations.  For the random-flux,  the curves depend on $\kp$.  We considered $L=30$ and average over $50$ link configurations for the random-flux case.}
	\label{fig:bottclean}
\end{figure}

As a benchmark,  we present results for the clean case in Fig.  \ref{fig:bottclean}.  For the $0$-flux state,  we get $\mathcal{B}=\pm 1$.  For the $1$-flux and $1/2$-flux cases,  see Fig.  \ref{fig:fluxes}, we have $\mathcal{B}=\pm 2$ as expected \citep{kitaev06,  takikawa20,  fuchs20}.  For the random-flux case, however,  the averaged value of $\mathcal{B}$ is no longer quantized,  even in the absence of disorder in the bonds,  $\delta K=0$. We interpret this result as the lack of a topological phase for this flux configuration.

To probe the extent of the bulk states, we employ a level statistics study of the spectrum \citep{prodan10}.  This investigation is instrumental in assessing the stability of the topological gap in the Majorana spectrum in the presence of disorder and complements the information coming from $\mathcal{B}$.  Specifically,  we study the average level spacing ratio \cite{oganesyan07}
\begin{align}\label{rstat}
\widetilde{r}_{\nu} = \frac{\mbox{min}(\delta_\nu, \delta_{\nu-1})}{\mbox{max}(\delta_\nu, \delta_{\nu-1})} = \mbox{min}\pa{\frac{r_\nu}{r_{\nu-1}}}.
\end{align}
Here $\delta_\nu = E_{\nu+1} - E_\nu$ is the difference between two adjacent energy levels $E_\nu$ from a given disorder realization and $r_\nu = \delta_\nu/\delta_{\nu-1}$. For extended states, the ratio follows the Gaussian unitary ensemble (GUE) statistics and the average value of $\widetilde{r}$ is $\mean{\widetilde{r}}_{\text{GUE}} \approx 0.60266$. On the other hand,  exponentially localized states follow the Poisson statistics, with $\mean{\widetilde{r}}_{\text{Poisson}} \approx 0.38629$ \citep{atas13}.   The definition in Eq.  \eqref{rstat} is particularly useful because it avoids the definition of a local average level spacing.  In practice,  we calculate $\mean{\widetilde{r}}$ at a given energy $E$ using a small energy window of $5$ levels above and below $E$ \citep{prodan10}.

\subsection{Extraction of the power-law exponent}

A vital result of the current work is the presence of a power-law DOS at low-energies.  To extract the power-law exponent,  we employ two complementary methods.  First,  we plot the curves $\rho(E) \sim E^{-\alpha}$ on a log-log scale and get $\alpha$ as the slope of the linear regression to the DOS curve.  As an alternative approach, we extract $\alpha$ directly from the spectrum \citep{clauset09}.
 
We assume that the energy histogram, DOS,  displays a power-law form in the interval $ 0 < E < E_{\rm{max}} $,  with $E_{\rm{max}}$ the upper cutoff energy below which the power-law holds. Given a data set containing $N_{\rm{ob}}$ observations $E \leq E_{\rm{max}}$,  the value of $\alpha$ for the power-law model that is most likely to have generated this data is given by \citep{clauset09}:
\begin{align}\label{estimator}
\alpha =  1 - \mean{ \ln\pa{ \frac{E_{\rm{max}}}{E} }}^{-1}_{E\leq E_{\text{max}}},
\end{align}
with the associated statistical: $\sigma_{\alpha} = \alpha/\sqrt{N_{\rm{ob}}}$. However, the greatest source of uncertainty is the definition of $E_{\rm{max}}$.  A simple way to find its optimal value is to plot $\alpha \times E_{\rm{max}}$,  picking $E_{\rm{max}}$ within an energy window where $\alpha$ is reasonably stable.  

\begin{figure}[t]
	\centering
	\includegraphics[width=1.0\linewidth]{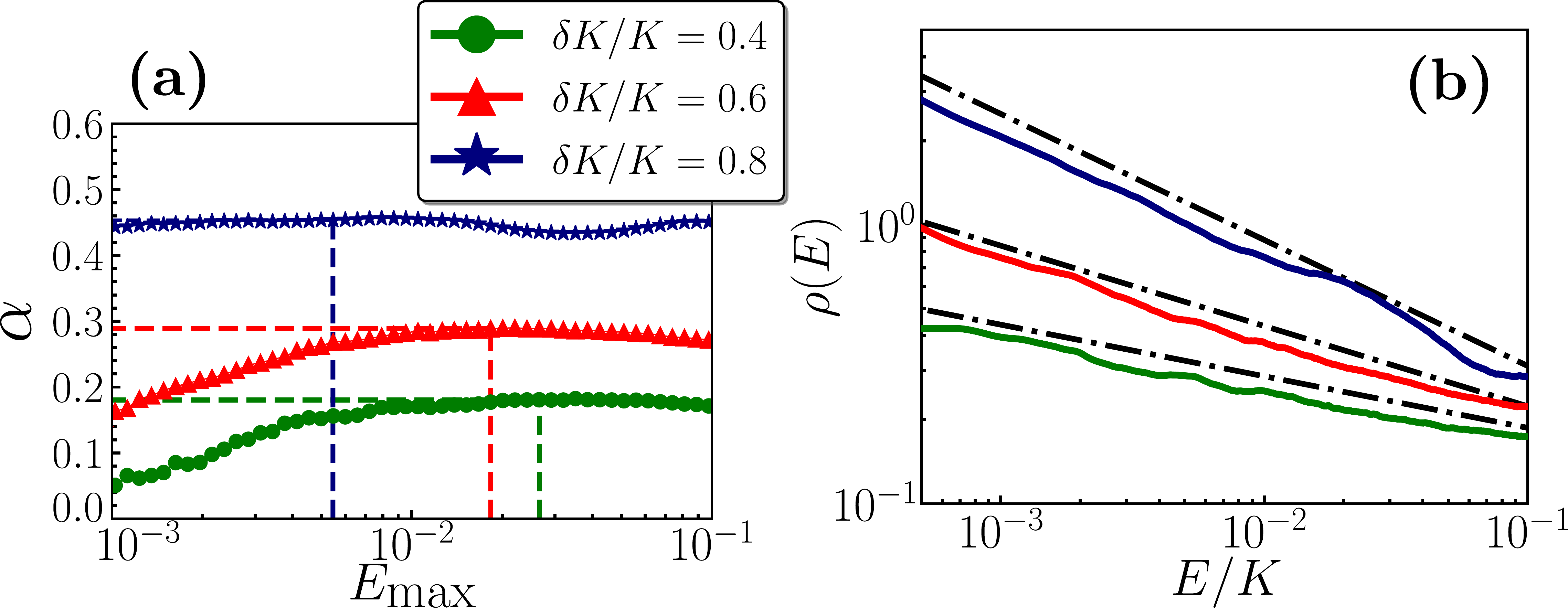}
	\caption{(a) Power-law exponent $\alpha$ as function of the upper cutoff energy below which the power-law holds $E_{\rm{max}}$.  We consider the random-flux configuration and bond disorder with $\kappa=\kp=0$.  The dashed lines indicate the optimal values of $\alpha$ and $E_{\rm{max}}$.  (b) DOS in a log-log scale at low energies.  The dot-dashed lines are power-law fits,  $\rho(E) \sim E^{-\alpha}$,  to the data with the exponent $\alpha$ extracted in (a).  The fits are shifted with respect to the DOS curves for the sake of clarity.  We considered $L=30$ and $3 \times 10^3$ realizations of disorder.}
	\label{fig:figs3}
\end{figure}

This method has proved to be quite satisfactory for our data.  As an example, we show the exponent extraction for the random-flux state with bond disorder,  Fig.\ref{fig:figs3}(a).  The values of $\alpha$ obtained from Eq.  \eqref{estimator} fit well the DOS at low energy as shown in Fig.\ref{fig:figs3}(b).

\section{Flux states}

\begin{figure}[t]
	\centering
	\includegraphics[width=1\linewidth]{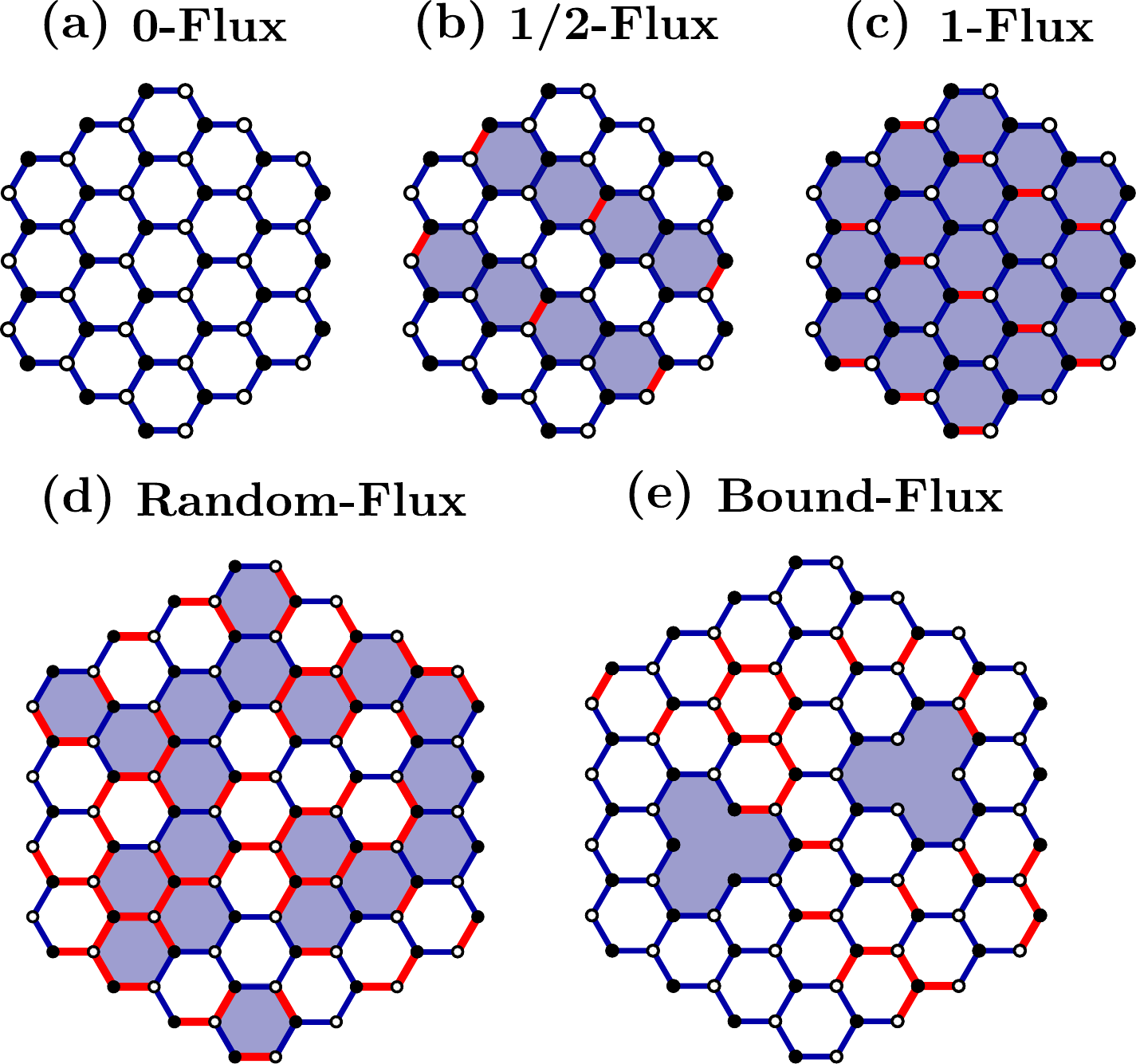}
	\caption{Flux configurations considered in this work.  Red (blue) bonds correspond to $u_{ij} = -1(+1)$,  and shaded (white) hexagons correspond to flux $W_p = -1\,(+1)$.  (a) $0$-flux link configuration,  (b) $1/2$-flux link configuration,  (c) $1$-flux link configuration,  (d) a random-flux link configuration,  and (e) a bound-flux link configuration. }
	\label{fig:fluxes}
\end{figure}

To perform the diagonalization of the free fermion Hamiltonian in Eq.  \eqref{eq:kitaev_majoranas}, we need to fix the link variables $u_{ij}$ and work in a well-defined static flux state.  We investigated three periodic flux configurations in our work: $0$-flux,  $W_{p}=+1$ for all hexagons [Fig. \ref{fig:fluxes}(a)],  $1/2$-flux,  $W_{p}=+1$ for half of the hexagons and $W_{p}=-1$ for the other half [Fig. \ref{fig:fluxes}(b)],  and $1$-flux,  $W_{p}=-1$ for all hexagons [Fig. \ref{fig:fluxes}(c)].  In the clean case,  the $0$-flux state is the ground state for $\kp=0$.  For $\kp \gtrsim 1/8$, the ground state corresponds to the $1$-flux state \citep{zhang19}.  The $1/2$-flux state is never a competitive ground state,  but it is instructive to study it since it serves as the periodic version of the random-flux state,  Fig. \ref{fig:fluxes}(d).  For the parameters we consider in this work,  we do not observe a transition in the flux state as a function of $\kappa$. 

In the random-flux state, we randomly assign $u_{ij}= \pm 1$ to each link with equal probability.  In average,  we have $\mean{W_{p}}=0$,  Fig. \ref{fig:fluxes}(d).   To construct the bound-flux configuration,  we follow the prescription from Ref.  \citep{kao21}.  After all vacancies positions are assigned,  we randomly flip a single bond around the $l=12$ site plaquette surrounding each of them,  see Fig.  \ref{2vac}(b).  This binds a flux inside this plaquette.  We then iteratively sweep over the lattice to guarantee that all hexagons not surrounding the defects encompass no flux.  We repeat this procedure for a given vacancy configuration -- starting at distinct random positions -- until we find a link configuration corresponding to a true bound-flux state. 

\begin{figure}[b]
	\centering
	\includegraphics[width=1.0\linewidth]{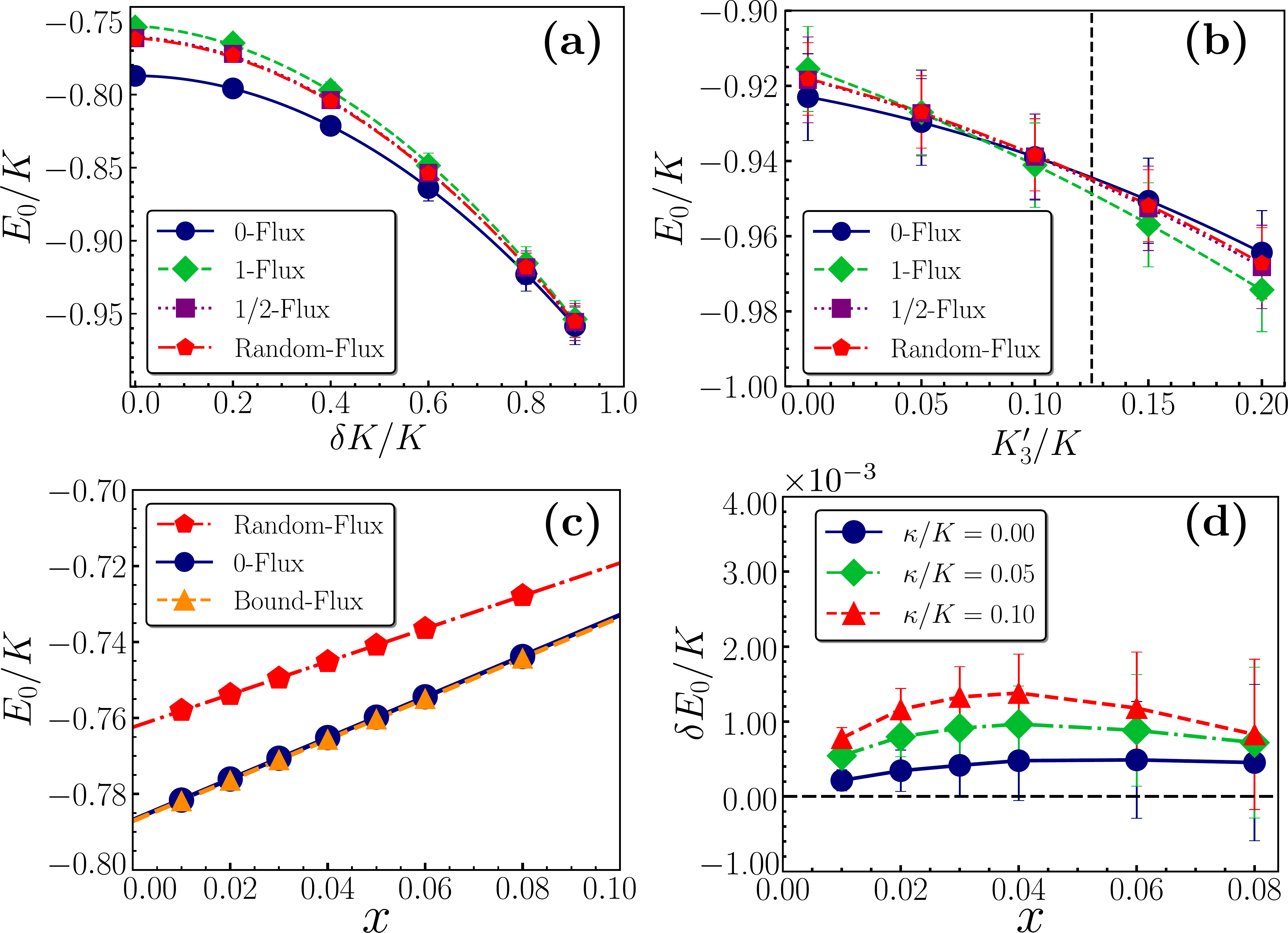}
	\caption{(a) Ground state energy per site,  $E_0$,  as a function of the bond disorder strength $\delta K$ for $\kappa=\kp=0$.  (b) $E_0$ as a function of $\kp$ for $\delta K=0.8$ and $\kappa=0$.  The vertical dashed line marks the position of the clean transition between the $0$-flux and $1$-flux state. (c) $E_0$ as a function of the vacancy concentration $x$ for $\kappa=\kp=0$.  (d) Difference between the ground state energy of the $0$-flux and bound-flux states,  $\delta E_0$ as a function of $x$ for $\kp=0$ and three values of $\kappa$.  We considered $L=30$ and up to $3 \times 10^3$ disorder realizations. }
	\label{ground}
\end{figure}

Once we define the static flux configuration,  we diagonalize the extended Kitaev model and compute its ground state energy.  The flux state with the smallest energy is selected as the ground state.  Although biased,  this variational approach is numerically efficient and exploits the integrability of the model.  The results of this procedure are illustrated in Fig.  \ref{ground}.  

In Fig.  \ref{ground}(a), we show the ground state energy as a function of the bond disorder for different flux configurations for $\kappa=\kp=0$.  At weak disorder,  the $0$-flux state is the ground state,  thus evolving adiabatically from clean limit \citep{kitaev06}.  All fluxes produce the same ground-state energy within error bars for $\delta K/K \gtrsim 0.6$.   We interpret this result as a tendency towards the random-flux state for strong disorder \citep{zschocke15}.  In Fig.  \ref{ground}(b) we show the ground state energy as a function of $\kp$ for fixed bond disorder $\delta K=0.8$.  All energies coincide within error bars.  Nevertheless,  there is a tendency towards the $1$-flux state as $\kp$ increases.  This is reminiscent of the transition in the clean limit at $\kp\approx1/8$. 

Fig.  \ref{ground}(c) shows the ground-state energy as a function of the vacancy concentration for $\kappa=\kp=0$.  The bound-flux and $0$-flux states have very similar energies,  with a slight preference for the bound-flux configuration.  The energy of the random-flux state is not competitive in this range of dilution.  In Fig.  \ref{ground}(d), we show the energy difference between the $0$-flux and bound-flux states as a function of $x$ for different $\kappa$ values and $\kp=0$.  There is no change in the ground state as a function of $\kappa$ for our parameter range,  suggesting the bound-flux as the ground state for $x \lesssim 0.1$.


\subsection{Rare-regions}

\begin{figure}[b]
	\centering
	\includegraphics[width=1.0\linewidth]{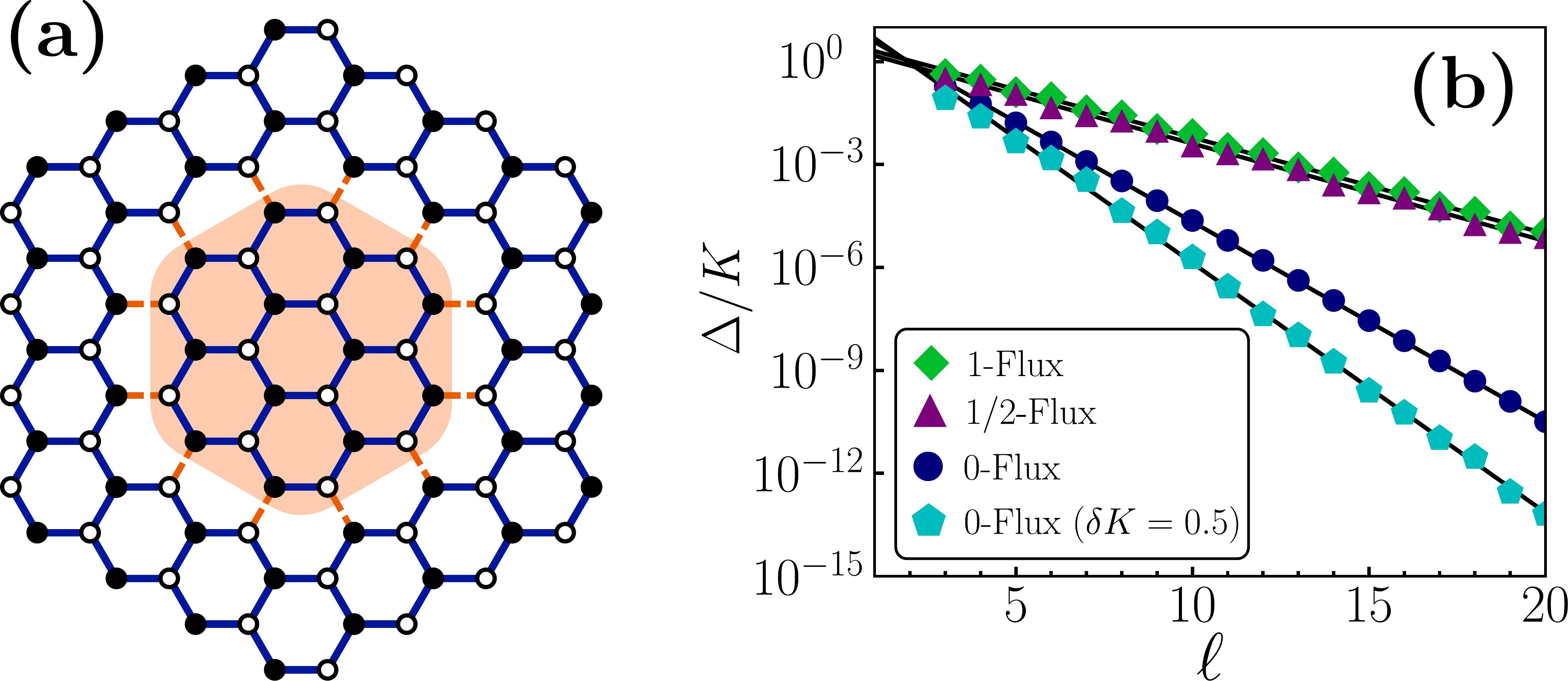}
	\caption{(a) Sketch of a disorder-induced rare-region (shaded area).  The dashed orange links represent the weak bonds connecting this droplet to the bulk.  (b) Finite-size energy gap $\Delta$ for the Kitaev model as a function of the cluster size $\ell$ considering open boundary conditions.  We consider different flux states in the clean case.  For the disordered case,  $\delta K=0.5$,  we show the result for the $0$-flux, and we average over $100$ realizations of disorder. }
	\label{rr}
\end{figure}

Fig.  \ref{rr}(a)  sketches a rare region: it is a droplet connected to the bulk through weak couplings only.  To estimate the scaling of the excitation gap with its linear size $\ell$,  we assume this island is completely disconnected from the bulk.  This simplification allows us to treat the rare regions as finite clusters of size $\ell$ and open boundary conditions.   This assumption is strictly valid for bond-disorder in the limit $\delta K \to K$, whereas it is readily realized for spin dilution.  Fig.  \ref{rr}(b) shows the scaling of the gap with $\ell$ in a mono-log scale, and the exponential decay is evident for all flux configurations we studied.  As discussed in the main text,  we interpret this exponential decay as a consequence of the hybridization between the (trivial) states localized at the edges of the open finite clusters. 

The pure Kitaev model displays a phase transition as a function of the exchange anisotropy \citep{kitaev06}.  Our work considers only isotropic Kitaev exchange couplings,  but disorder renders them locally anisotropic.  For instance,  if a given site has two weak and one strong bond, it could be locally inside the gapped phase of the pure Kitaev model if $\delta K > K/3$.  It is then natural to ask if taking into account disorder inside the rare region would modify our argument qualitatively.  We find that this is not the case,  with the finite-size energy gap still going exponentially to zero as the size of the rare region is increased,  Fig.   \ref{rr}(b).  This result points towards the robustness of our argument.  We leave an in-depth study of the transition between the two phases in the Kitaev model in the presence of disorder for future work.

\section{Further results for bond disorder}

\begin{figure}[t]
	\centering
	\includegraphics[width=0.9\linewidth]{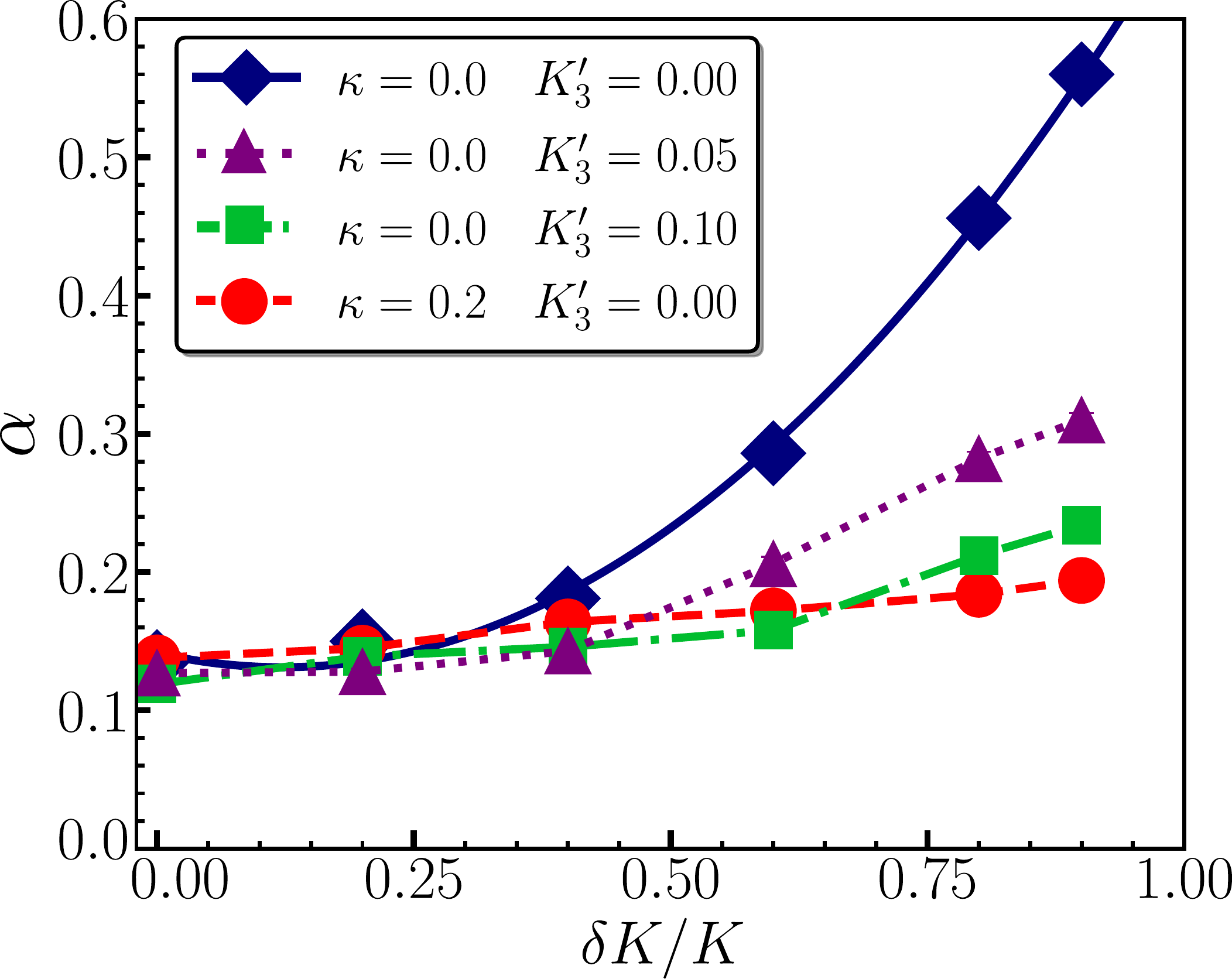}
	\caption{Non-universal power-law exponent $\alpha$ for the extended Kitaev model,  Eq.  \eqref{eq:kitaev_majoranas},  in the random-flux state with bond disorder as a function of the disorder strength $\delta K$.  We observe $\alpha=0.13$ for $\delta K=0$.  We consider different values of $\kappa$ and $\kp$,  $L=30$,  and $3 \times 10^3$ disorder realizations.}
	\label{alpha_bond}
\end{figure}

We present further results for the extended Kitaev model in Eq.  \eqref{eq:kitaev_majoranas} in the presence of bond disorder.  In Fig.  \ref{alpha_bond} we show the non-universal power-law exponent $\alpha$ as a function of $\delta K/K$ and different values of $\kappa$ and $\kp$.  Considering the random-flux configuration,  we get $\alpha=0.13$ for $\delta K = \kappa = \kp= 0$,  highlighting the fact that the random-flux favors an accumulation of low-$E$ states even in the absence of disorder. We see that $\alpha$ increases with disorder, and it is suppressed by $\kappa$ and $\kp$.  As $\delta K$ increases,  the coupling of the rare regions to the bulk is weakened. One may translate this effect into enhancing the effective probability of finding a droplet of size $\ell$, which accounts for the enhancement of $\alpha$.  In the Majorana language,  $\kappa$ and $\kp$ correspond to second and third-neighbor hopping.  Because these longer-range hopping amplitudes are not disordered,  they effectively increase the coupling of the rare regions to the bulk and thus decrease $\alpha$.  This is qualitatively similar to the effect of long-range magnetic interactions in the usual Griffiths scenario \citep{miranda05,  vojta10}. 

For \hlio, it was reported that $\alpha=1/2$ \cite{kitagawa18}.  The Griffiths-like scenario we propose implies that thermodynamics and local dynamics alone are insufficient to conclusively pin down the minimum model for this Kitaev material since $\alpha$ is non-universal.  Further crucial information for \hlio, and other Kitaev materials,  comes from their topological properties.  

For our model,  with bond disorder alone,  the power-law behavior in $\rho(E)$ is robust only for the random-flux state.  The power-law disappears for the $0$-flux state as we switch on $\kappa$ because the clean topological gap survives.  This automatically implies that the topological phase of the Kitaev model is stable for bond disorder, provided the flux configuration of the clean limit is preserved,  which is valid for weak to moderate disorder. 

\begin{figure}[t]
	\centering
	\includegraphics[width=1\linewidth]{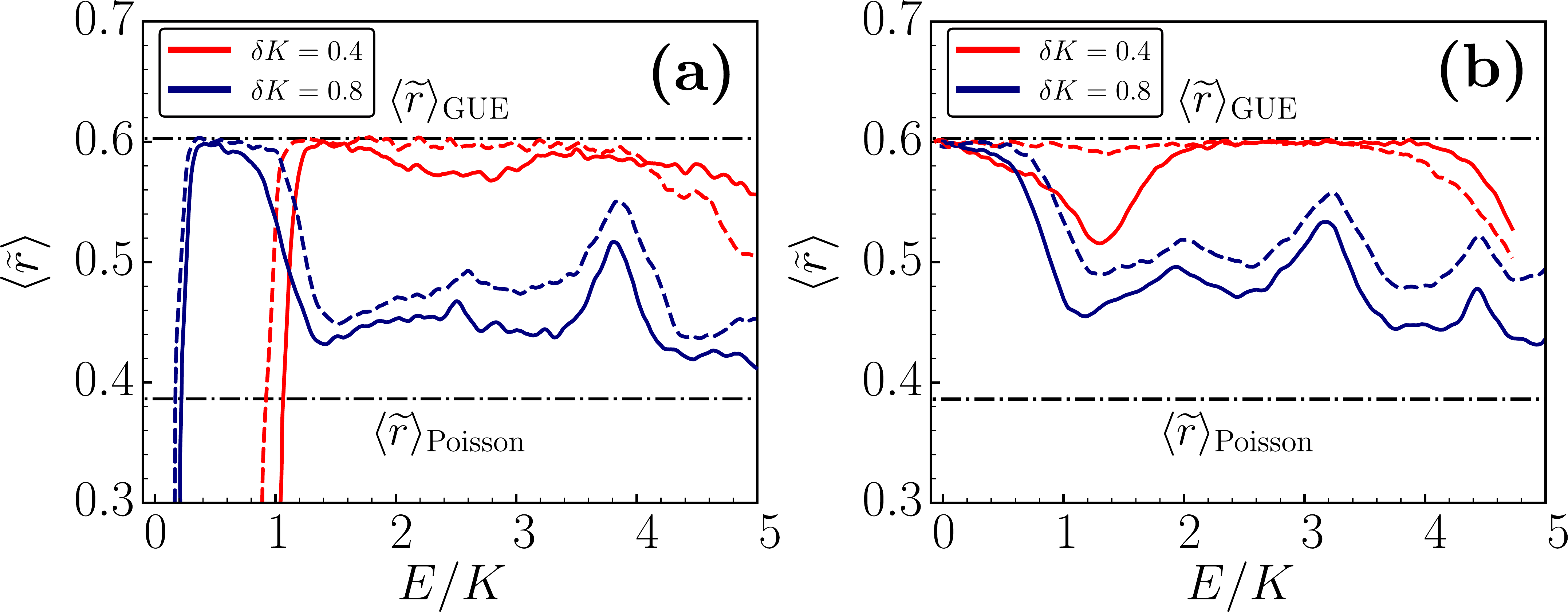}
	\caption{Average level spacing ratio for the extended Kitaev model,  Eq.  \eqref{eq:kitaev_majoranas},  with bond disorder as a function of the energy $E$ for $\kappa =0.2K$ and different values of $\delta K$ and $\kp$.  The full (dashed) curves correspond to $\kp=0\,(0.1)$.  (a) $0$-flux state.  (b) Random-flux state.  We considered $L=30$ and $3 \times 10^3$ realizations of disorder.}
	\label{levelbond}
\end{figure}

The robustness of the topological phase in our model reflects not only in a quantized Bott index but also in the level spacing statistics,  Fig.  \ref{levelbond}.  For the $0$-flux, we see that $\mean{\widetilde{r}}$ touches the expected value for GUE statistics just before the topological gap then drops abruptly inside it due to the absence of states in this region.  While putative extended states for higher energies,  $E \gtrsim 2K$,  are suppressed with the disorder,  the extended states close to the gap edge are remarkably robust.  We ascribe the shrinking of the topological gap with $\delta K$ to the levitation and annihilation mechanism first identified in disordered Chern insulators \citep{onoda07}.  For the random-flux state,  the topological gap does not exist. There is no evidence of extended states,  except at $E=0$,  which is a particular point in the model,  even for this flux state,  due to the symmetric bond disorder we consider.

\section{Further results for vacancies}

\begin{figure}[t]
	\centering
	\includegraphics[width=1\linewidth]{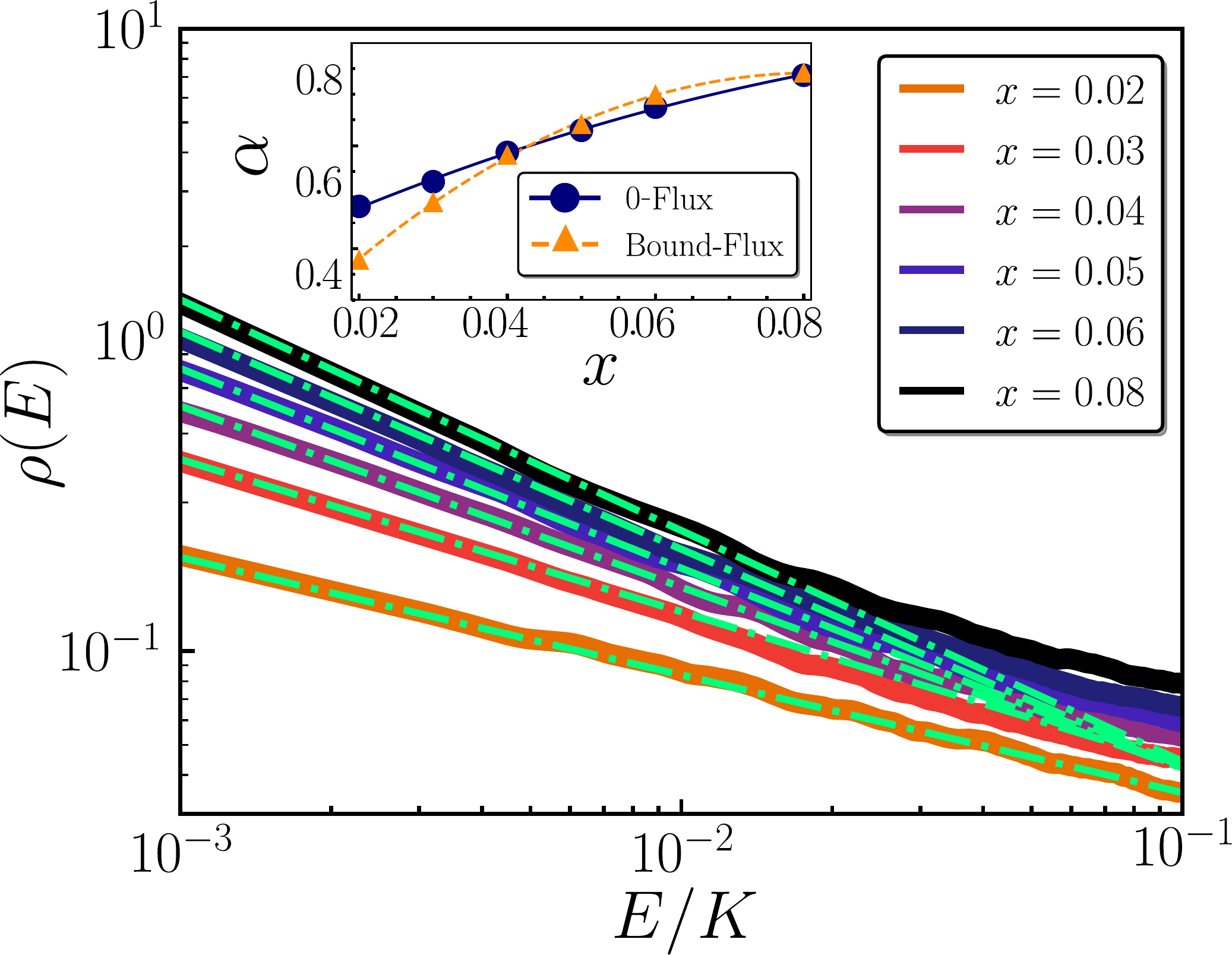}
	\caption{ Low-energy behavior of the density of states in a log-log plot for several values of vacancy concentration $x$ in the bound-flux state.  The dot-dashed lines are power-law fits to the data.  Inset: Non-universal power-law exponent $\alpha$  as a function of  $x$ both for the $0$-flux and bound-flux configurations.   We consider $\kappa=\kp=0$,  $L=30$,  and $10^3$ disorder realizations.}
	\label{alpha_vac}
\end{figure}

We now present different results for the extended Kitaev model in Eq.  \eqref{eq:kitaev_majoranas} in the presence of vacancies.  In the inset of Fig.  \ref{alpha_vac}, we show $\alpha$ as a function of the vacancy concentration $x$ for both the bound-flux and the $0$-flux state.  The behavior in both cases is similar to the one in Fig.  \ref{alpha_bond}:  $\alpha$ increases with $x$ because the larger the dilution,  the larger the odds of constructing a droplet of linear size $\ell$ disconnected from the bulk.  Similarly,  we find that $\alpha$ is suppressed with longer range hopping.  In Fig.  \ref{alpha_vac} we show the low-energy part of the DOS, highlighting its power-law behavior for the bound-flux configuration.  The curves for the $0$-flux configuration are similar.

\begin{figure}[b]
	\centering
	\includegraphics[width=1\linewidth]{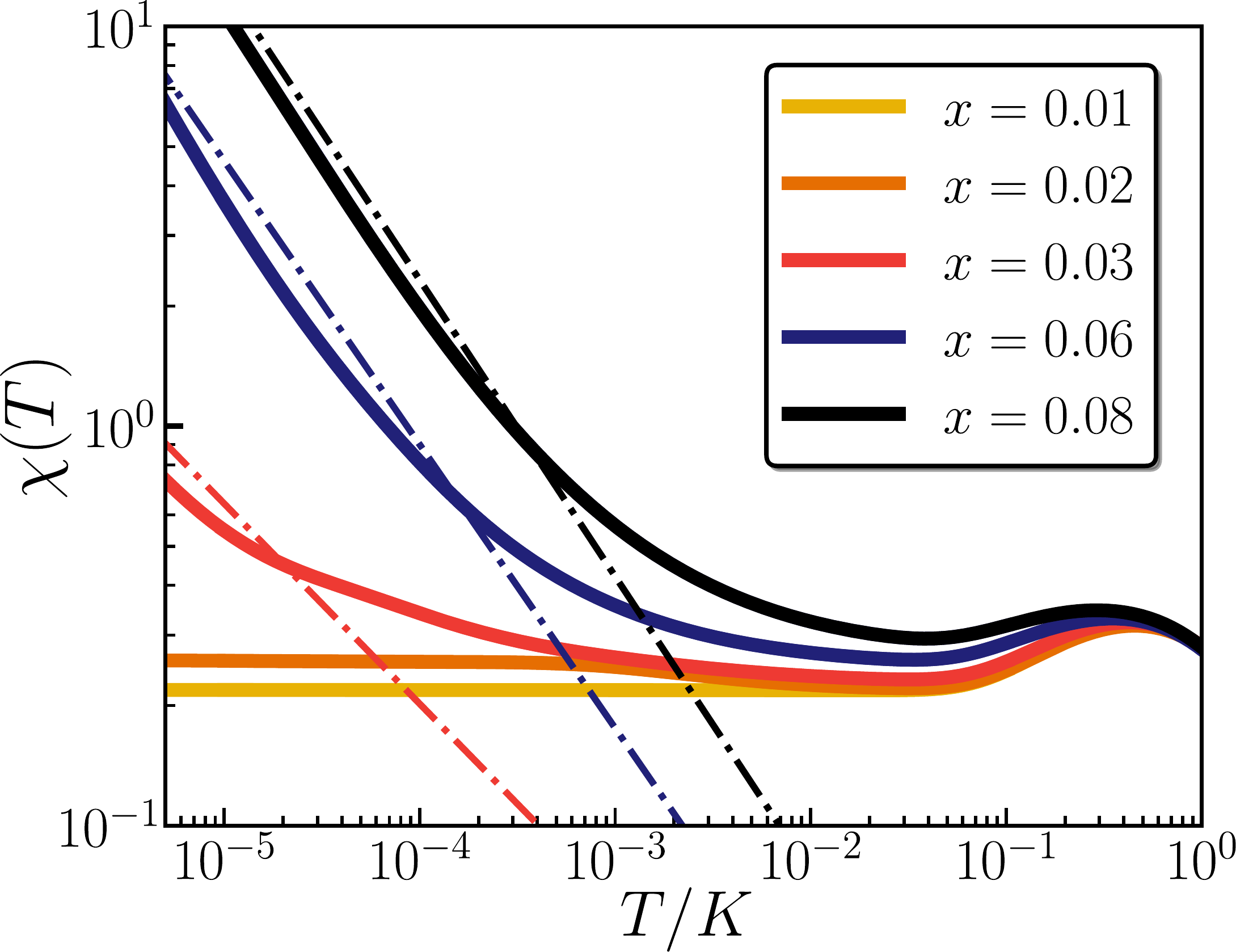}
	\caption{ Low-temperature part of the uniform susceptibility $\chi(T)$ as a function of the temperature $T$ in a log-log plot for several values of vacancy concentration $x$ in the bound-flux state.  The dot-dashed lines are power-law fits to the data,  with power coming from Fig.  \ref{alpha_vac}.   We consider $\kappa=\kp=0$,  $L=30$,  and $10^3$ disorder realizations.}
	\label{chi_vac}
\end{figure}

This power-law divergence of the DOS at low energies also manifests itself in the physical observables. In Fig.  \ref{chi_vac}, we show the uniform susceptibility for the bound-flux state,  which behaves as $\chi(T) \sim T^{-\alpha}$ at very low-$T$.  The contribution for this power-law tail comes from the unpaired spins because $\Delta_{2f}=0$ for these sites,  Fig.  \ref{fig:figs1}(b).  Since the density of unpaired spins goes as $x$,  the singular behavior becomes more pronounced at very low-$T$  and larger dilution.  In these regimes,  it overcomes the regular contribution of the bulk spins.  Experimentally,  this should translate into a milder divergence of the uniform susceptibility,  at not too low $T$ and small $x$,  compared to the specific heat. The results for the $0$-flux state are again similar.

As in the case of bond disorder,  a more apparent distinction between the different static fluxes emerges for $\kappa \neq 0$.  The Bott index is no longer quantized for the $0$-flux state if  $x > 0.02$ in the small $\kappa$ regime.  Interestingly,  it remains pinned to an integer value for larger values of $x$ in the bound-flux state,  although the clean topological gap is the same.  The loss of quantization in both fluxes comes from the emergence of an impurity band inside the clean topological gap,  similar to what is observed in disordered Chern insulators \citep{onoda07,  prodan10,  castro15}. 

\begin{figure}[t]
	\centering
	\includegraphics[width=1\linewidth]{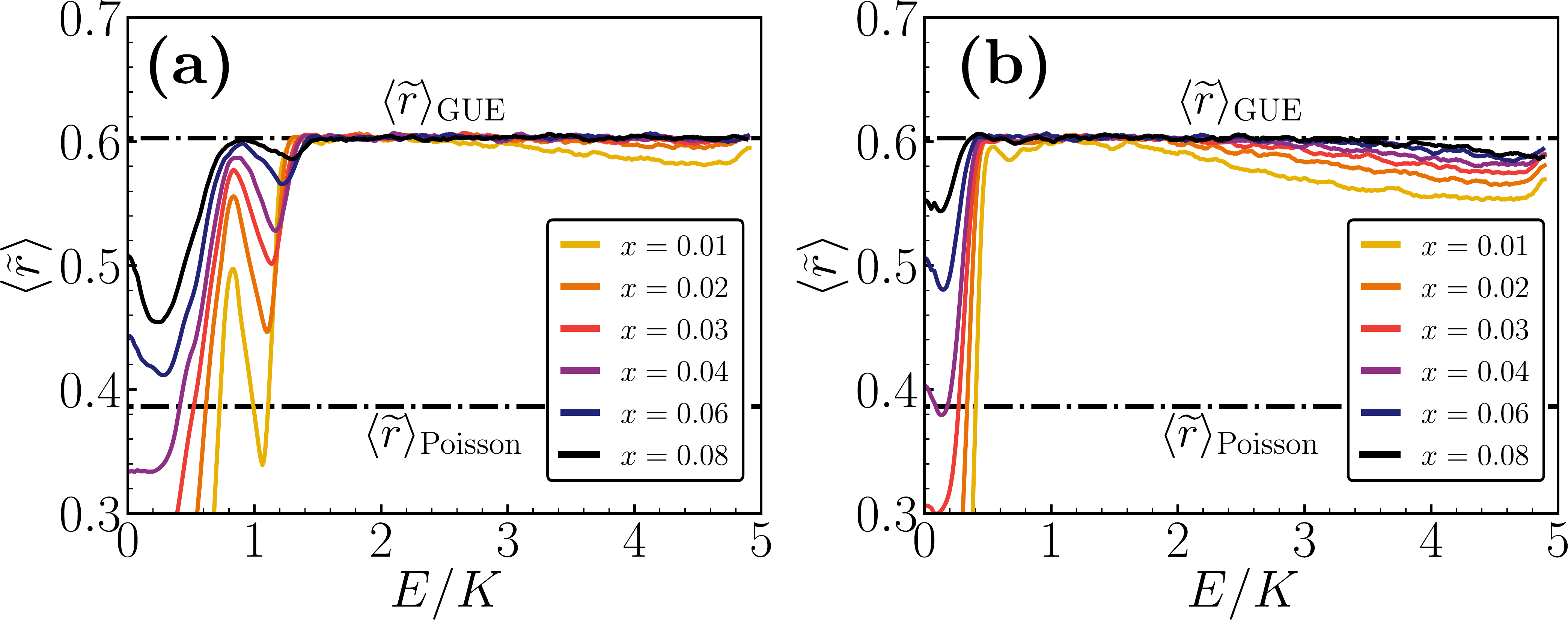}
	\caption{ Average level spacing ratio in the bound-flux state as a function of the energy $E$ for $\kp=0.1K$ and several concentrations $x$.  (a) $\kappa=0.1K$.  (b) $\kappa=0.05K$.  We considered $L=30$ and $3 \times 10^3$ realizations of disorder.}
	\label{levelvac}
\end{figure}

This extra robustness for the topological phase in the bound-flux state comes from the fact that each vacancy binds a flux \citep{willans10}, which creates a localized level inside the clean gap. The larger the $\kappa$,  the more well-defined this state, and the topological phase survives for larger $x$.  The averaged level spacing ratio for the bound-flux case is shown in Fig.  \ref{levelvac}.  For $\kappa=0.05K$ and $\kp=0.1K$,  the energy of the aforementioned localized state is close to the edge of the clean topological gap, and the topological phase survives only up to $x\approx0.03$.  For $\kappa=\kp=0.1K$,  there is a well-defined in-gap state at $E \approx 0.8K $.  As the impurity levels move initially into this localized state,  there is extra protection, and the topological survives up to $x \approx 0.05$.

\begin{figure}[b]
	\centering
	\includegraphics[width=1\linewidth]{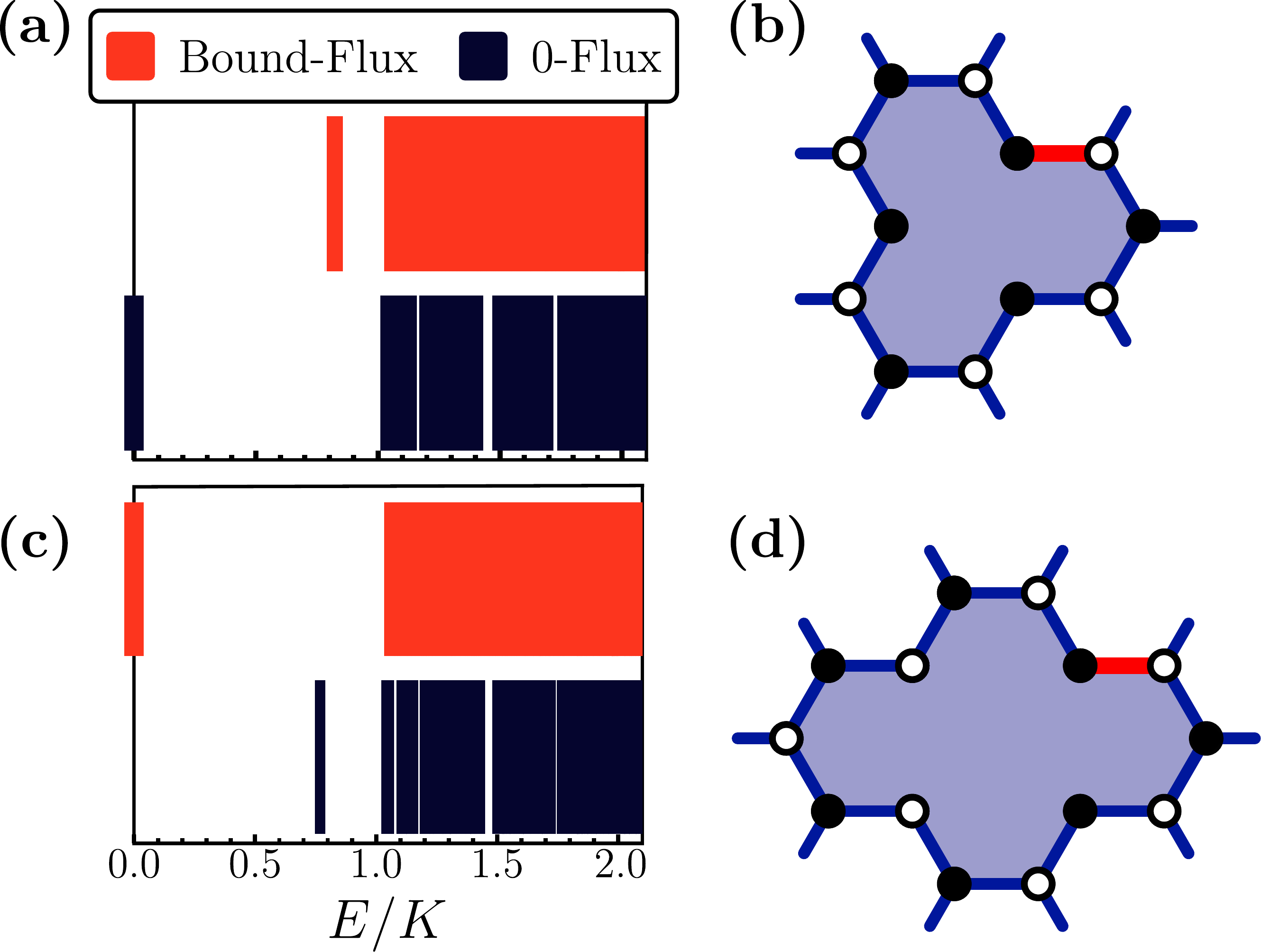}
	\caption{(a) Energy levels for a system with two vacancies placed at a distance $L/2$ apart.  We show the spectrum for the $0$-flux and bound-flux states.  (b) $l=12$ plaquette corresponding to a single vacancy.  (c) Energy levels for a system with two pairs of neighboring vacancies placed at a distance $L/2$ apart.  We show the spectrum for the $0$-flux and bound-flux.  (d) $l=14$ plaquette corresponding to a pair of neighboring vacancies.  We considered $\kappa=0.1K$ and $\kp=0.0$. }
	\label{2vac}
\end{figure}

To understand the appearance of this in-gap state at finite energies, it is sufficient to study the problem of two vacancies placed in an otherwise $0$-flux clean background.  We set these impurities at a distance $L/2$ apart,  but our results are independent of this distance as long as the impurities plaquettes do not share a common link \citep{kao21}.  An impurity plaquette with $l=12$ sites is shown in Fig.  \ref{2vac}(b).  We consider two situations: the impurity either binds a flux (bound-flux) or does not ($0$-flux).  We then diagonalize Eq.  \eqref{eq:kitaev_majoranas} and study its energy spectrum,  Fig.  \ref{2vac}(a).  For the $0$-flux state,  there is an $E=0$ state (a Majorana zero mode), whereas the impurity energy is gapped for the bound-flux state.  This is the source of extra protection for the topological phase in the bound-flux state.

Interestingly,  the gap for the two impurities problem gives precisely the energy of the in-gap state for our full numerics.  The existence of an energy gap in the bound-flux state can be traced back to an even more straightforward setup.   Take a $l=12$ tight-binding chain with nearest-neighbor hopping only.  The spectrum of this problem has (has not) a gap if the chain binds (does not bind) a flux.  

As $x$ increases,  the diluted impurity picture breaks down.  In particular,  the probability of finding a pair of neighboring vacancies becomes non-negligible.  In this situation, the length of the two-impurity plaquette is $l=14$,  Fig.  \ref{2vac}(d).  Suppose we study the energy levels of two pairs of neighboring impurities separated by a distance $L/2$. In that case, the result is reversed with respect to the single vacancy case: the bound-flux state displays a Majorana zero mode,  whereas the $0$-flux state shows a gap,  Fig.  \ref{2vac}(c).  Ultimately,  the topological phase is destroyed with the increase of dilution for all static flux states, and we observe $\rho(E) \sim E^{-\alpha}$ at low-$E$.  Nevertheless,  the topological phase is particularly robust for the bound-flux state at small $x$, and increasing $\kappa$ helps stabilize it.


%